\documentclass[twocolumn]{aastex63}

\shorttitle{WASP43~b Spitzer}
\shortauthors{Morello G. et al.}

\usepackage{amsmath,nicefrac}
\usepackage{enumitem}
\usepackage{multirow}
\usepackage[flushleft]{threeparttable}
\usepackage{stackengine}
\usepackage{tabularx}

\begin{document}

\defcitealias{stevenson17}{S17}
\defcitealias{mendonca18}{M18}


\title{An independent analysis of the \textit{Spitzer}/IRAC phase curves of WASP43~b}


\author{G. Morello, C. Danielski, D. Dickens}
\affil{AIM, CEA, CNRS, Universit\'e Paris-Saclay, Universit\'e Paris Diderot, Sorbonne Paris Cit\'e, F-91191 Gif-sur-Yvette, France}
\email{giuseppe.morello@cea.fr}
\author{P. Tremblin}
\affil{Maison de la Simulation, CEA, CNRS, Univ. Paris-Sud, UVSQ, Universit\'e Paris-Saclay, 91191 Gif-sur-Yvette, France}
\author{ P.-O. Lagage}
\affil{AIM, CEA, CNRS, Universit\'e Paris-Saclay, Universit\'e Paris Diderot, Sorbonne Paris Cit\'e, F-91191 Gif-sur-Yvette, France}



\begin{abstract}
We present here a reanalysis of the \textit{Spitzer Space Telescope} phase curves 
of the hot Jupiter WASP43~b, using the wavelet pixel-Independent Component Analysis, a blind signal-source separation method.
The data analyzed were recorded with the InfraRed Array Camera and consisted of two visits at 3.6~$\mu$m, and one visit at 4.5~$\mu$m, each visit covering one transit and two eclipse events.
To test the robustness of our technique we repeated the analysis on smaller portions of the phase curves, and by employing different instrument ramp models.
Our reanalysis presents significant updates of the planetary parameters compared to those reported in the original phase curve study of WASP43~b. In particular, we found (1) higher nightside temperatures, (2) smaller hotspot offsets, (3) a greater consistency ($\sim$1~$\sigma$) between the two 3.6~$\mu$m visits, and (4) a greater similarity with the predictions of the atmospheric circulation models. Our parameter results are consistent within 1~$\sigma$ with those reported by a recent reanalysis of the same data sets.
For each visit we studied the variation of the retrieved transit parameters as a function of various sets of stellar limb-darkening coefficients, finding significant degeneracy between the limb-darkening models and the analysis output.
Furthermore, we performed the analysis of the single transit and eclipse events, and we examined the differences between these results with the ones obtained with the whole phase curve.
Finally we provide a formula useful to optimize the trade-off between precision and duration of observations of transiting exoplanets.
\end{abstract}





\section{INTRODUCTION}

WASP43~b is a hot Jupiter orbiting around a K7~V star in $\sim$19.5~hours \citep{hellier11}. Table~\ref{tab1} reports the stellar, planetary and transit parameters taken from the WASP43~b discovery paper \citep{hellier11}. The ultra-short orbital period of WASP43~b has inspired multiple observational programs of its full phase curve using the \textit{Hubble Space Telescope} (\textit{HST}, \citealp{stevenson14}), and the \textit{Spitzer Space Telescope} (\citealp{stevenson17}, hereinafter \citetalias{stevenson17}). Another full phase curve observation of WASP43~b is planned as part of the Transiting Exoplanet Community Early Release Science program of the \textit{James Webb Space Telescope} using the Mid-InfraRed Instrument \citep{bean18}.

Exoplanet phase curves are measurements of the flux coming from a star+exoplanet system as a function of the orbital phase. If the exoplanet is transiting, its phase curve includes (usually) both transit and eclipse events. The flux modulations observed in the mid-infrared are attributed to the thermal emission from the exoplanet with varying phase angle \citep{cooper05, fortney06, cowan07}.
Exoplanets with short orbital periods are expected to be tidally locked to their host star \citep{showman02}, therefore exhibiting a hotter dayside and a cooler nightside. The day--night temperature contrast depends on the heat recirculation efficiency of the exoplanetary atmosphere. Numerical simulations also predict a (model-dependent) hotspot offset from the substellar point \citep{showman02, cooper05, kataria15, schwartz17, zhang17}.

\begin{table}[!t]
\begin{center}
\begin{threeparttable}
\caption{WASP43 system parameters \label{tab1}}
\begin{tabular}{lc}
\tableline\tableline
\multicolumn{2}{c}{Stellar parameters} \\
\tableline
$T_{\mbox{\footnotesize eff}}$ (K) & 4400$\pm$200 \\
$\log{g_*}$ (cgs) & 4.65$_{-0.04}^{+0.06}$ \\
$[\mbox{Fe/H}]$ (dex) & -0.05$\pm$0.17 \\
$M_*$ ($M_{\odot}$) & 0.58$\pm$0.05 \\
$R_*$ ($R_{\odot}$) & 0.60$_{-0.04}^{+0.03}$ \\
\tableline
\multicolumn{2}{c}{Planetary parameters} \\
\tableline
$M_p$ ($M_{\mbox{\footnotesize Jup}}$) & 1.78$\pm$0.10 \\
$R_p$ ($R_{\mbox{\footnotesize Jup}}$) & 0.93$_{-0.09}^{+0.07}$ \\
$a$ (au) & 0.0142$\pm$0.0004 \\
\tableline
\multicolumn{2}{c}{Transit parameters} \\
\tableline
$p^2$ & 0.0255$\pm$0.0012 \\
$b$ & 0.66$_{-0.07}^{+0.04}$ \\
$i$ (deg) & 82.6$_{-0.9}^{+1.3}$ \\
$P$ (days) & 0.813475$\pm$0.000001 \\
$E.T.$ (HJD) & 2455528.86774$\pm$0.00014 \\
\tableline
\end{tabular}
\begin{tablenotes}
\item From \cite{hellier11}.
\end{tablenotes}
\end{threeparttable}
\end{center}
\end{table}

\cite{stevenson14} and \citetalias{stevenson17} claimed extremely low circulation efficiency for the atmosphere of WASP43~b: $\varepsilon =$0.002$_{-0.002}^{+0.01}$, where $\varepsilon$ is the night--day bolometric flux ratio. They also detected a wavelength-dependent eastward hotspot offset, i.e., their phase curve models peak prior to secondary eclipses. However, \citetalias{stevenson17} discarded the first 3.6~$\mu$m data set, which presented discrepant results, and larger correlated noise in the light curve residuals. \citetalias{stevenson17} also discarded a $\sim$2 hr interval from the second 3.6~$\mu$m data set, corresponding to an unexpected flux decrement in their detrendend light curve, that the authors attributed to unmodeled instrumental or astrophysical red noise.

The low nightside fluxes and the large eastward offsets measured by \citetalias{stevenson17} in the \textit{Spitzer}/InfraRed Array Camera (IRAC) passbands could not be reproduced by using the SPARC/MITgcm code of \cite{kataria15}. The SPARC/MITgcm is a 3D global circulation model coupled to a non-gray radiative transfer code. \cite{keating17} pointed out that the atmosphere of WASP43~b should have a much higher circulation efficiency, $\varepsilon \sim$0.5, based on the inverse correlation between the day-night temperature contrast and stellar irradiation \citep{cowan11, perez-becker13, schwartz15, komacek16}. 

\cite{mendonca18}, hereinafter \citetalias{mendonca18}, reanalyzed the three \textit{Spitzer}/IRAC phase curves without discarding any data. \citetalias{mendonca18} found a better agreement between the two 3.6~$\mu$m observations, and higher nightside fluxes than those reported by \citetalias{stevenson17}.

In this paper we present an independent reanalysis of the three \textit{Spitzer}/IRAC phase curves of WASP43~b using the wavelet pixel-Independent Component Analysis (ICA) pipeline \citep{morello16}. We repeated the analysis by adopting different stellar limb-darkening models, which affect the retrieved transit parameters. We compare our results with those reported by \citetalias{stevenson17} and \citetalias{mendonca18}, and with theoretical expectations. In addition to the full phase curve analyses, we explore the ability to constrain the different parameters from shorter observations, nominally half phase curves, transit-only, and eclipse-only. This kind of study will be useful for planning future JWST proposals, and optimizing the time schedule of the Atmospheric Remote-sensing Infrared Exoplanet Large-survey (ARIEL) mission, in order to maximize their scientific return.

\begin{table*}[!ht]
\caption{\textit{Spitzer}/IRAC data sets analyzed for this study. \label{tab2}}
\begin{tabular*}{\textwidth}{c @{\extracolsep{\fill}}cccccccc}
\tableline\tableline
Obs.\tablenotemark{a} &  Prog. ID & AORs\tablenotemark{b} & UT Date & $\Delta$t (h)\tablenotemark{c} & Mode\tablenotemark{d} & Pip. \tablenotemark{e} \\
\tableline
Ch1, visit 1  & 11001 & 52364544 & 2015 Mar 7 & 8.5 & sub, 2.0 & 19.2.0 \\
(3.6 $\mu$m) &  & 52364800 & 2015 Mar 7 & 8.5 & sub, 2.0 & 19.2.0 \\
 &  & 52355312 & 2015 Mar 8 & 8.5 & sub, 2.0 & 19.2.0 \\
\tableline
Ch1, visit 2  & 11001 & 57744384 & 2015 Sep 5 & 15.2 & sub, 2.0 & 19.2.0 \\
(3.6 $\mu$m) &  & 57744640 & 2015 Sep 5 & 10.2 & sub, 2.0 & 19.2.0 \\
\tableline
Ch2  & 10169 & 51777024 & 2014 Aug 27 & 8.5 & sub, 2.0 & 19.2.0 \\
(4.5 $\mu$m) &  & 51777280 & 2014 Aug 27 & 8.5 & sub, 2.0 & 19.2.0 \\
 &  & 51777792 & 2014 Aug 28 & 8.5 & sub, 2.0 & 19.2.0 \\
\tableline
\end{tabular*}
\tablenotetext{a}{IRAC channel, visit number, and wavelength.}
\tablenotetext{b}{Astronomical Observation Requests.}
\tablenotetext{c}{Total duration of the AOR in hours.}
\tablenotetext{d}{Readout mode and frame time in seconds.}
\tablenotetext{e}{Pipeline version of the Basic Calibrated Data.}
\end{table*}

\section{OBSERVATIONS}
\label{sec:observations}
We reanalyzed three \textit{Spitzer}/IRAC observations of the phase curve of WASP43~b. Each visit consists of two to three consecutive Astronomical Observation Requests (AORs) over a 25.4~hr interval, including one transit and two eclipse events. Observational and detector information for the individual data sets is given in Table~\ref{tab2}.

\section{DATA ANALYSIS}
\label{sec:analysis}

\begin{table*}[!ht]
\begin{threeparttable}
\caption{Claret-4 limb-darkening coefficients for the WASP43 star in the 3.6  and 4.5~$\mu$m \textit{Spitzer}/IRAC passbands. \label{tab_ldc}}
\begin{tabular*}{\textwidth}{c @{\extracolsep{\fill}}cccccc}
\tableline\tableline
Method & Channel  & $a_1$ & $a_2$ & $a_3$ & $a_4$ \\
\tableline
\multirow{2}{*}{A17} & Ch1, 3.6 $\mu$m & 0.596193 & -0.353618 & 0.234039 & -0.070725 \\
 & Ch2, 4.5 $\mu$m & 0.574190 & -0.585735 & 0.550643 & -0.199545 \\
\tableline
\multirow{2}{*}{A100} & Ch1, 3.6 $\mu$m & 0.575555 & -0.288784 & 0.154642 & -0.037594 \\
 & Ch2, 4.5 $\mu$m & 0.538245 & -0.472568 & 0.411592 & -0.141342 \\
\tableline
\multirow{2}{*}{P100} & Ch1, 3.6 $\mu$m & 4.843472 & -10.282954 & 10.828015 & -4.187545 \\
 & Ch2, 4.5 $\mu$m & 4.846365 & -10.290074 & 10.835657 & -4.190506 \\
\tableline
\multirow{2}{*}{PQS} & Ch1, 3.6 $\mu$m & 0.763637 & 0.265362 & -0.458262 & 0.090083 \\
 & Ch2, 4.5 $\mu$m & 0.763610 & 0.265746 & -0.458614 & 0.090143 \\
\tableline
\end{tabular*}
\begin{tablenotes}
\item Using ATLAS9 and PHOENIX stellar-atmosphere models; calculated with the code by \cite{espinoza15}, available at http://www.github.com/nespinoza/limb-darkening/.
\end{tablenotes}
\end{threeparttable}
\end{table*}

\begin{figure*}[!t]
\epsscale{0.95}
\plotone{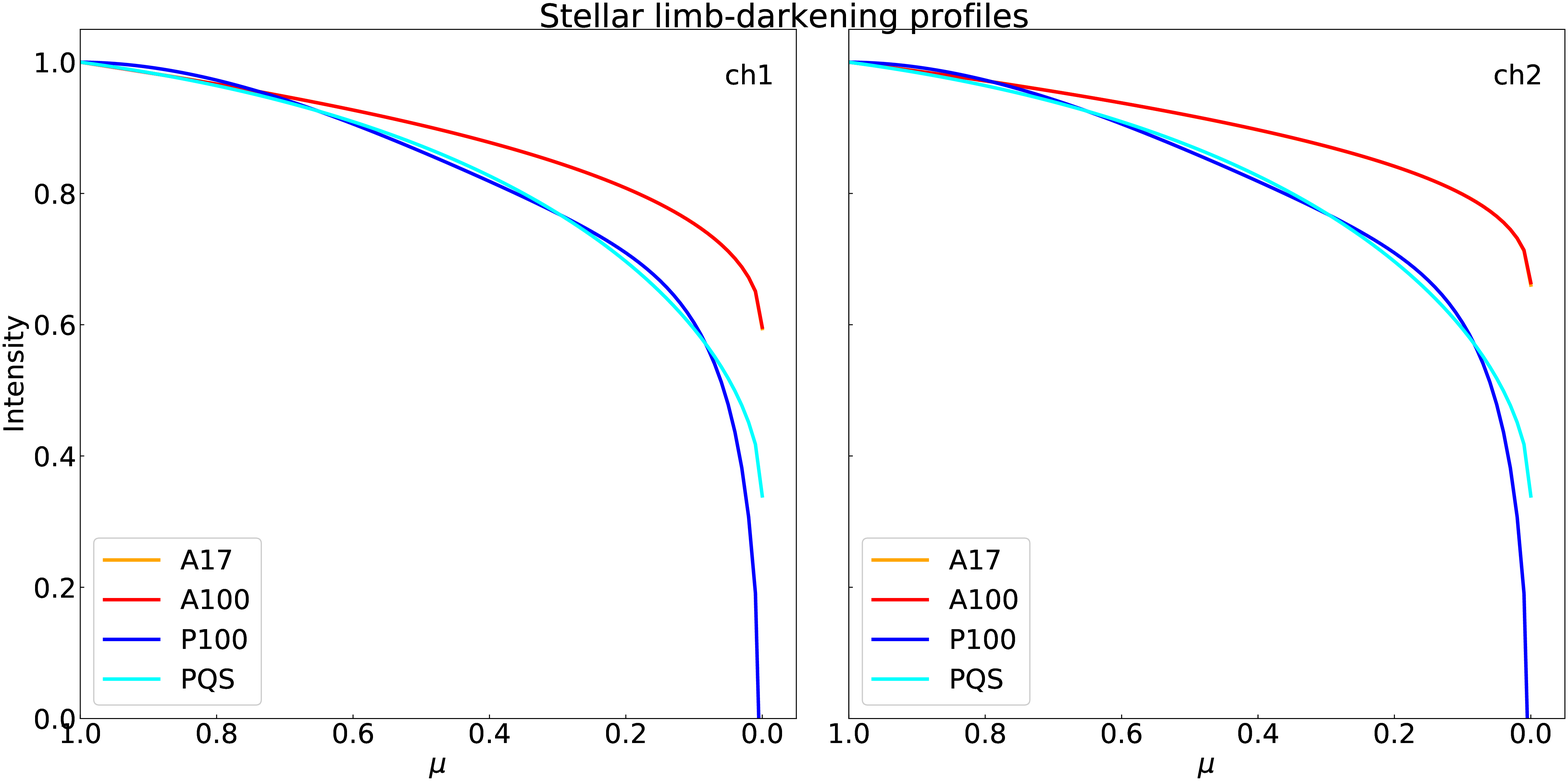}
\caption{Left panel: stellar limb-darkening profiles of the WASP43 star in the 3.6~$\mu$m \textit{Spitzer}/IRAC channel, computed by using the code provided by \cite{espinoza15} at http://www.github.com/nespinoza/limb-darkening/, with different settings: A17 (orange), A100 (red), P100 (blue), and PQS (cyan). Right panel: analogous plot for the 4.5~$\mu$m \textit{Spitzer}/IRAC channel.
\label{fig:ld_profiles}}
\end{figure*}

\subsection{The phase curve model}
\label{ssec:pc_model}
In our model the stellar flux is constant in time, and normalized to 1. The exoplanetary flux is given by
\begin{equation}
\label{eqn:pc_stev}
c_0 + c_1 \cos{ [2 \pi ( \Phi - \Delta \Phi - c_2)] } + c_3 \cos{ [4 \pi ( \Phi - \Delta \Phi - c_4)] } ,
\end{equation}
where $\Phi$ is the so-called orbital phase, i.e., the time from the reference epoch of transit ($E.T.$) in units of the orbital period ($P$), $\Delta \Phi$ is the mid-transit phase offset, and $c_0$--$c_4$ are free parameters used to model the phase curve modulations.  Equation~\ref{eqn:pc_stev} is equivalent to the formula adopted by \citetalias{stevenson17} and \citetalias{mendonca18}.
We used the formalism of \cite{mandel02} for modeling the exoplanetary transit and eclipses.

\subsection{Stellar limb-darkening coefficients}
\label{ssec:limb-darkening}
We calculated multiple sets of four-coefficient limb-darkening \citep{claret00}, hereinafter claret-4, for the WASP43 star in the 3.6 and 4.5~$\mu$m \textit{Spitzer}/IRAC passbands, using the code provided by \cite{espinoza15} at GitHub\footnote{http://www.github.com/nespinoza/limb-darkening/}. The code adopts two grids of stellar-atmosphere intensity models, i.e., ATLAS9\footnote{http://kurucz.harvard.edu/grids.html} \citep{kurucz79} and PHOENIX \citep{husser13}. The intensities in the models are given as a function of $\mu = \cos{\theta}$, where $\theta$ is the angle between the surface normal and the line of sight. The ATLAS models adopt a plane-parallel approximation for the stellar atmosphere, while the PHOENIX models use spherical geometry. As a consequence, the PHOENIX models show a characteristic steep drop-off in intensity at small, but finite $\mu$ values, which is not well approximated by any of the standard parametric laws \citep{claret12, claret13, morello17}. The limb-darkening coefficients also depend on the sampling of the intensities \citep{howarth11, neilson13, neilson13b, espinoza15}. We tested the following fitting options:
\begin{itemize}
\item A17, i.e., least-squares fit to the ATLAS model intensities calculated at 17 angles;
\item A100, i.e., least-squares fit to the ATLAS intensities interpolated at 100 angles, uniformly sampled in $\mu$, with a cubic spline;
\item P100, i.e., least-squares fit to the PHOENIX intensities interpolated at 100 angles, uniformly sampled in $\mu$, with a cubic spline;
\item PQS, i.e., least-squares fit to the PHOENIX model intensities with $\mu \ge$0.1 (quasi-spherical models, as defined by \citealp{claret12}).
\end{itemize}
We discarded the least-squares fit to all the PHOENIX model intensities, because it led to anomalous (non-monotonic) limb-darkening profiles. The most likely cause of the anomalous results was that the PHOENIX model intensities are more finely sampled near the steep drop-off, which is then overweighted in the fit.
We interpolated the limb-darkening coefficients in $T_{\mbox{\footnotesize eff}}$ and $\log{g}$ to the WASP43 parameter values reported in Table~\ref{tab1}. Table~\ref{tab_ldc} reports the four sets of claret-4 limb-darkening coefficients obtained with the different fitting options.
Figure~\ref{fig:ld_profiles} shows the corresponding intensity profiles. We note that the ATLAS limb-darkening profiles, A17 and A100, overlap in the plot. The PHOENIX profiles, P100 and PQS, indicate stronger limb-darkening than the ATLAS profiles. The P100 profiles reach zero intensity at the stellar limb, while the PQS profiles remain significantly above zero. Note that the PQS profiles are not accurate at the stellar limb, given that their behavior is extrapolated from the model intensities with $\mu \ge$0.1.

\begin{figure*}[!t]
\epsscale{0.95}
\plotone{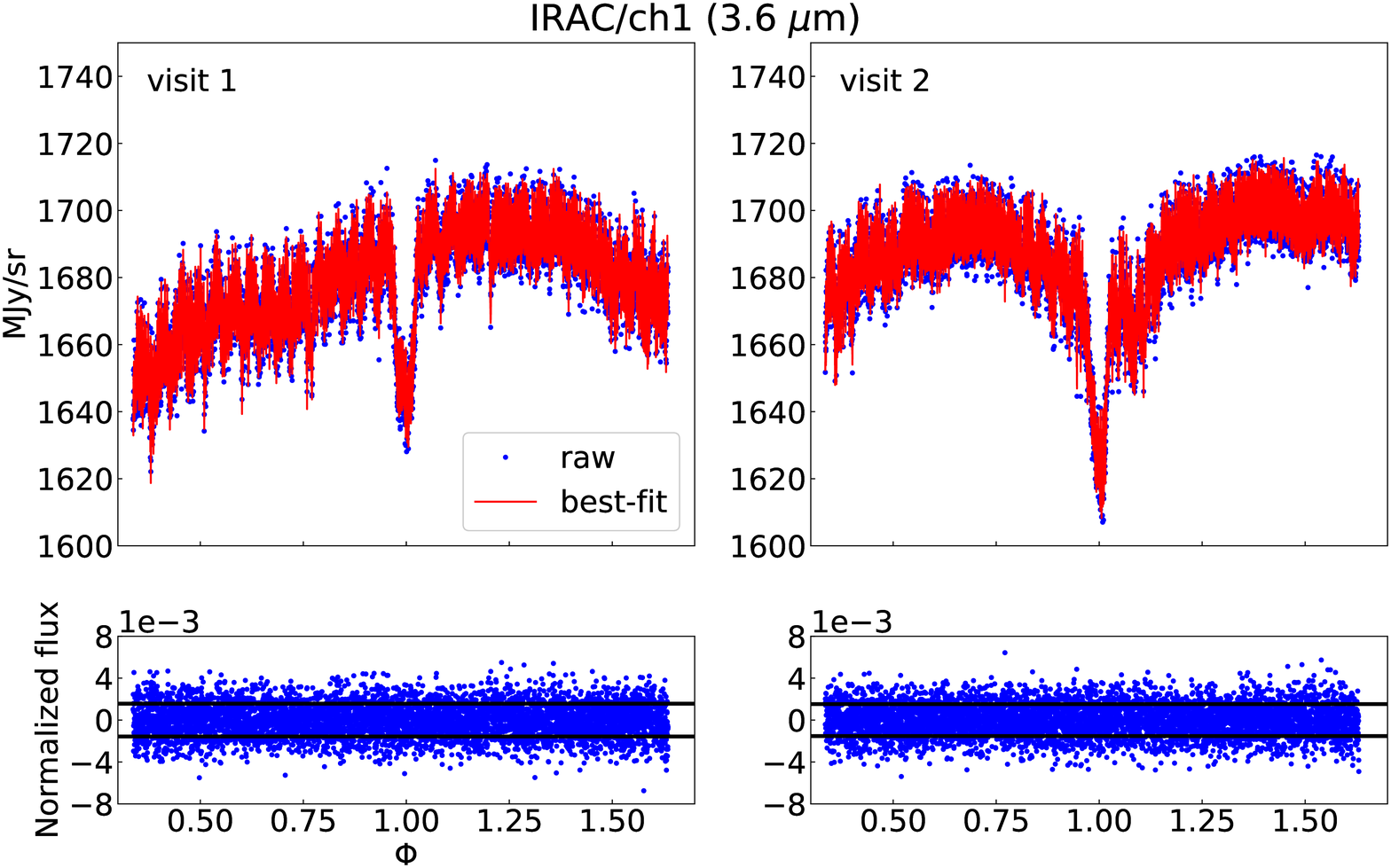}
\caption{Top panels: raw light curves (blue dots) obtained for the \textit{Spitzer}/IRAC observations at 3.6 $\mu$m, and relevant best-fit models (red line). Bottom panels: residuals from the above light curves and models (blue points), and standard deviations (black lines).
\label{fig:rawbest_ch1}}
\end{figure*}
\begin{figure}[!h]
\epsscale{0.99}
\plotone{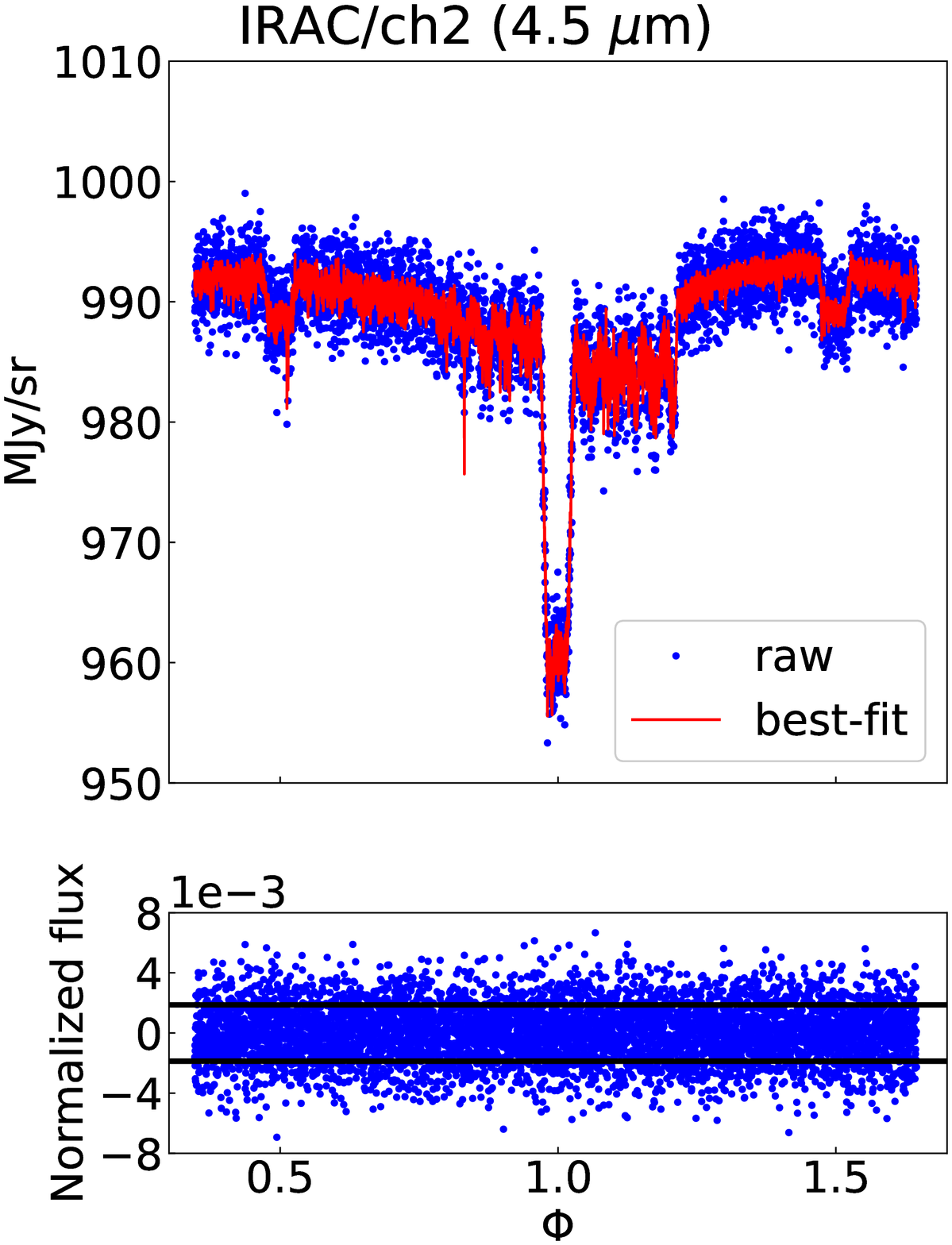}
\caption{Top panel: raw light curve (blue dots) obtained for the \textit{Spitzer}/IRAC observations at 4.5 $\mu$m, and relevant best-fit model (red line). Bottom panel: residuals from the above light curve and model (blue points), and standard deviations (black lines).
\label{fig:rawbest_ch2}}
\end{figure}

\subsection{Detrending \textit{Spitzer}/IRAC data}
For our analysis we used the Basic Calibrated Data (BCD) provided by the \emph{Spitzer} Heritage Archive \citep{wu10}. BCD are flat-fielded, and flux-calibrated frames (\citealp{fazio04, irac_handbook}). We extracted the individual pixel time series from a 5$\times$5 array having the stellar centroid at its center, and computed the sum-of-pixel time series, here referred to as raw light curves. We binned the time series by a factor of 8, i.e., temporal bin size of 16~s, in order to reduce the computational time for the data analysis. The chosen bin size is smaller than the time scales of interest, e.g., it is $\sim$1/63 of the transit ingress duration.

Then, we applied the wavelet pixel-ICA technique \citep{morello16} to simultaneously fit the phase curve model and the instrumental effects. We also tested the time pixel-ICA technique \citep{morello15b}, obtaining similar or less robust results which we report in appendix~\ref{app_wavetime}, together with a detailed comparison of both techniques.
Both algorithms rely on ICA, i.e., a blind source separation technique, to extract the instrumental components from the light curves. Such blind approaches have proven to perform as well as or better than other state-of-the-art pipelines to detrend \textit{Spitzer}/IRAC observations of exoplanetary transits and eclipses \citep{morello15, ingalls16}.

In this work, the pixel-ICA pipelines were applied to full phase curve observations, which may be affected by detector systematics with longer time scales compared to the transit-only and eclipse-only observations. We checked for residual long-trend systematics by adding a linear or quadratic function of time in the light curve fits, and by comparing the differences in the Bayesian Information Criterion (BIC, \citealp{schwarz78}) obtained with these various ramp models (constant, linear, or quadratic), as suggested by \citetalias{stevenson17}. Then, following the Occam's Razor principle, we confirmed the solution obtained with the constant ramp, if it had the lowest BIC. In an alternative case, the model selection was based on a number of considerations that will be explained in the following sections.

\section{RESULTS}
\label{sec:results}

The BIC favored the pure ``wavelet ICA + phase curve'' (constant ramp) models for the 4.5~$\mu$m and first 3.6~$\mu$m visits. For the second 3.6~$\mu$m visit the lowest BIC was obtained with the quadratic ramp model, while the BIC obtained with the constant ramp model was the highest (see Table~\ref{tabBICAIC}). We observed that the best-fit astrophysical parameters do not significantly depend on the choice of the ramp parameterization, except for the phase curve parameters of the second 3.6~$\mu$m visit (see Section~\ref{ssec:pc_results} and Appendix~\ref{app_halfpc}).

\begin{figure}[!t]
\epsscale{0.99}
\plotone{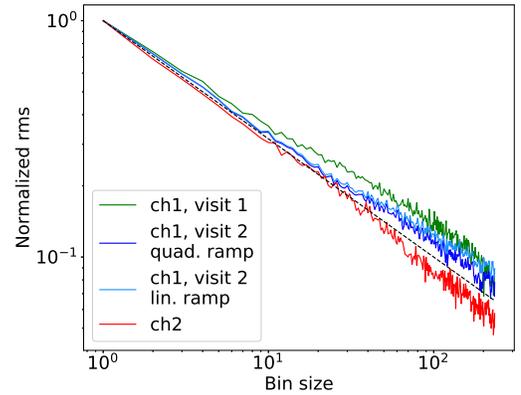}
\caption{Normalized rms of residuals as function of bin size for the first 3.6~$\mu$m visit (green), second 3.6~$\mu$m visit using a quadratic (blue) or linear (dodger blue) ramp model, and 4.5~$\mu$m visit (red). The black dashed line shows the theoretical behavior for gaussian residuals.
\label{fig:binning_rmsres_scale_ch1e2_best}}
\end{figure}

Figures~\ref{fig:rawbest_ch1} and \ref{fig:rawbest_ch2} show the raw light curves, the relevant best-fit models (with the minimum BIC), and the residuals. The rms amplitudes of the normalized residuals are 1.56$\times$10$^{-3}$ for the first 3.6~$\mu$m visit, 1.52$\times$10$^{-3}$ for the second 3.6~$\mu$m visit, and 1.87$\times$10$^{-3}$ for the 4.5~$\mu$m visit. We estimate them to be $\sim$24$\%$, 22$\%$, and 4$\%$ above the photon noise limit. Figure~\ref{fig:binning_rmsres_scale_ch1e2_best} shows how the rms amplitudes of the fitting residuals scale as a function of the bin size. The 4.5~$\mu$m residuals show no significant deviations from the theoretical behavior of white noise, different from the 3.6~$\mu$m residuals. The amount of residual correlated noise in the second 3.6~$\mu$m visit is notably smaller than in the first visit.

\begin{figure*}[!t]
\epsscale{0.95}
\plotone{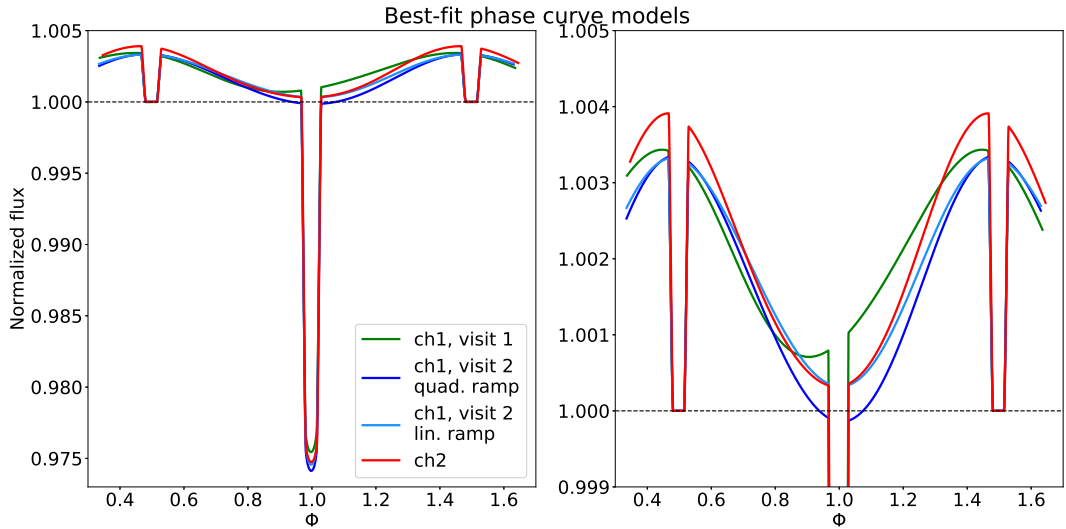}
\caption{Left panel: best-fit phase curve models for the first 3.6~$\mu$m visit (green), second 3.6~$\mu$m visit using a quadratic (blue) or linear (dodger blue) ramp model, and 4.5~$\mu$m visit (red). The black horizontal line indicates the stellar flux level. Right panel: zoom-in of the left panel.
\label{fig:phasecurves_best}}
\end{figure*}

\begin{figure*}[!t]
\epsscale{0.75}
\plotone{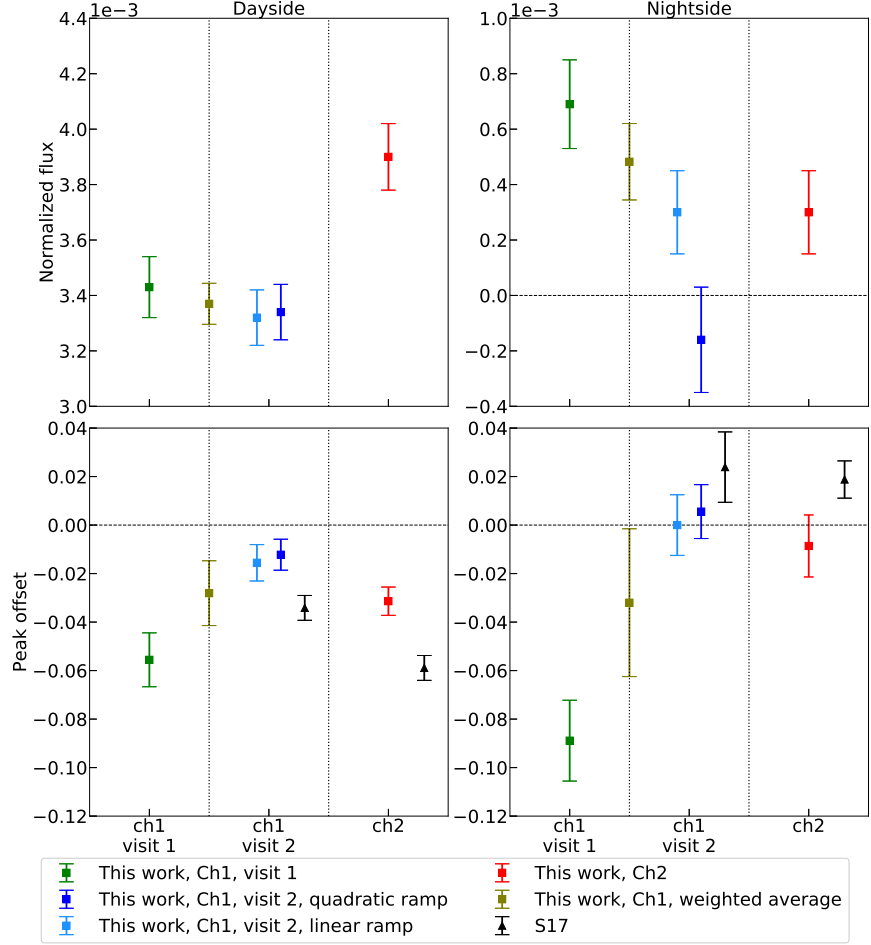}
\caption{Left, top panel: maximum exoplanetary flux relative to the stellar flux, for the first 3.6~$\mu$m visit (green square), second 3.6~$\mu$m visit using a quadratic (blue) or linear (dodger blue) ramp model, weighted average between the first and second visit with a linear ramp (olive), and 4.5~$\mu$m visit (red square). Left, bottom panel: orbital phase of the maximum exoplanetary flux relative to mid-eclipse, including the values reported by \citetalias{stevenson17} (black triangles). Right panels: analogous plots for the minimum exoplanetary to the stellar flux.
\label{fig:daynight_best_withcomparstev}}
\end{figure*}

\subsection{Phase curve models and parameters}
\label{ssec:pc_results}

Figure~\ref{fig:phasecurves_best} shows the best-fit phase curve models.  Figure~\ref{fig:daynight_best_withcomparstev} reports the corresponding estimates of the planet dayside maximum and nightside minimum flux normalized to the stellar flux ($F_{\mbox{\footnotesize day}}^{\mbox{\footnotesize MAX}}$ and $F_{\mbox{\footnotesize night}}^{\footnotesize MIN}$), and their offsets relative to the mid-eclipse and mid-transit times ($\Delta \Phi_{\mbox{\footnotesize day}}^{\mbox{\footnotesize MAX}}$ and $\Delta \Phi_{\mbox{\footnotesize night}}^{\footnotesize MIN}$) respectively.

The results obtained for the second 3.6~$\mu$m data set with the quadratic ramp parameterization appear to be unphysical, yielding negative nightside fluxes, but still consistent with zero at the 1~$\sigma$ level. The results obtained with the linear ramp parameterization are more plausible, because they can be explained by a simpler physical model (see Section~\ref{ssec:EDIBT}). We present here the two sets of results for the second 3.6~$\mu$m data set, together with the selected results for the other data sets. A more detailed discussion about the model selection criteria is reported in Section~\ref{ssec:model_selection}.

The (normalized) planet dayside flux at 4.5~$\mu$m is (3.90$\pm$0.12)$\times$10$^{-3}$. The nightside flux is (3.0$\pm$1.5)$\times$10$^{-4}$. The maximum dayside flux occurs 37$\pm$7 minutes prior to the mid-eclipse time, which corresponds to a shift of 11.3$^\circ \pm$2.1$^\circ$ east of the substellar point. The minimum nightside flux occurs within the interval -10$\pm$15 minutes relative to the mid-transit time, i.e., between 7.2$^\circ$ east and 1.5$^\circ$ west of the anti-stellar point.

The 3.6~$\mu$m phase curve models have remarkably different amplitudes and shapes, but with similar dayside fluxes: (3.43$\pm$0.11)$\times$10$^{-3}$ for the first visit, and (3.34$\pm$0.10)$\times$10$^{-3}$ (quadratic ramp) or (3.32$\pm$0.10)$\times$10$^{-3}$ (linear ramp) for the second visit. The three estimates are mutually consistent within 0.5~$\sigma$, and smaller than the 4.5~$\mu$m maximum with 2--2.5~$\sigma$ significance level. The planet flux minima are (6.9$\pm$1.6)$\times$10$^{-4}$ for the first 3.6~$\mu$m visit, and (-1.6$\pm$1.9)$\times$10$^{-4}$ (quadratic ramp) or (3.0$\pm$1.5)$\times$10$^{-4}$ (linear ramp) for the second visit.

For the second 3.6~$\mu$m visit, the phase curve maximum occurs 14$\pm$7 minutes (quadratic ramp), or 18$\pm$9 minutes (linear ramp), earlier than the mid-eclipse time. These offsets correspond to hotspot shifts of 4.4$^\circ \pm$2.3$^\circ$ and 5.6$^\circ \pm$2.7$^\circ$ east of the substellar point. The phase curve minimum occurs at +6$\pm$14 minutes (quadratic ramp), or 0$_{-14}^{+16}$ minutes (linear ramp), relative to the mid-transit time. These offsets correspond to shifts of 2$^\circ\pm$4$^\circ$ and 0$_{-4}^{+5}$$^\circ$ west of the anti-stellar point. 

The first 3.6~$\mu$m phase curve model is strongly asymmetric, with peaks occurring 64$\pm$13 (maximum) and 103$\pm$18 (minimum) minutes earlier than the mid-eclipse and mid-transit time, or, equivalently, 20$\pm$4$^\circ$ and 32$^\circ \pm$6$^\circ$ East of the substellar and anti-stellar points.

However, the tests reported in the Appendix~\ref{app_halfpc} suggest that the true uncertainties in the peak offsets estimated for the 3.6~$\mu$m observations may be larger than the nominal error bars. For example, our estimate of the dayside peak offset for the first 3.6~$\mu$m visit becomes 18$\pm$9 minutes before mid-eclipse when considering only the first two out of three AORs, which is identical to the estimate from the second visit (linear ramp). The corresponding dayside and nightside fluxes are consistent with those obtained from the full data set analysis within 1~$\sigma$. The same tests confirm the robustness of the parameter estimates for the 4.5~$\mu$m observation within their nominal error bars.

\begin{table}[!t]
\begin{center}
\begin{threeparttable}
\caption{Phase curve parameters of WASP43~b. \label{tab_pc_final}}
\begin{tabular}{ccc}
\tableline\tableline
Parameter & 3.6~$\mu$m & 4.5~$\mu$m \\
\tableline
$F_{\mbox{\footnotesize day}}^{\mbox{\footnotesize MAX}}$ & (3.37$\pm$0.07)$\times$10$^{-3}$ & (3.90$\pm$0.12)$\times$10$^{-3}$ \\
$F_{\mbox{\footnotesize night}}^{\mbox{\footnotesize MIN}}$ & (4.8$\pm$1.4)$\times$10$^{-4}$ & (3.0$\pm$1.5)$\times$10$^{-4}$ \\
$\Delta \Phi_{\mbox{\footnotesize day}}^{\mbox{\footnotesize MAX}}$ & -0.028$\pm$0.013 & -0.031$\pm$0.006 \\
$\Delta \Phi_{\mbox{\footnotesize night}}^{\mbox{\footnotesize MIN}}$ & -0.032$\pm$0.030 & -0.009$\pm$0.013 \\
\tableline
\end{tabular}
\begin{tablenotes}
\item The 3.6~$\mu$m parameters are the weigthed mean values over the two visits (using the linear ramp for the second visit); the error bars on $F_{\mbox{\footnotesize night}}^{\mbox{\footnotesize MIN}}$, $\Delta \Phi_{\mbox{\footnotesize day}}^{\mbox{\footnotesize MAX}}$, and $\Delta \Phi_{\mbox{\footnotesize night}}^{\mbox{\footnotesize MIN}}$ are inflated by the difference between the individual estimates in units of $\sigma$ (factors of 1.26, 2.15, and 3.05, respectively).
\end{tablenotes}
\end{threeparttable}
\end{center}
\end{table}

Table~\ref{tab_pc_final} reports our final measurements of the day and nightside fluxes and peak offsets at 3.6 and 4.5~$\mu$m. The results at 3.6~$\mu$m are the weighted averages between those obtained for the two visits, with inflated error bars for those parameters which were not consistent within 1~$\sigma$. We discarded the (unphyisical) results obtained for the second 3.6~$\mu$m visit with a quadratic ramp, for reasons that will be further elaborated in Sections~\ref{ssec:model_selection} and \ref{ssec:EDIBT}.

\begin{table*}[!t]
\begin{center}
\begin{threeparttable}
\caption{Mean values of the transit parameters. \label{tab_means}}
\begin{tabular*}{\textwidth}{c @{\extracolsep{\fill}}cccccc}
\tableline\tableline
L-D & Mean & $b$ & $T_0$ (s) & $p^2\times$10$^{-2}$ (3.6 $\mu$m) & $p^2\times$10$^{-2}$ (4.5 $\mu$m) \\
\tableline
\multirow{2}{*}{A17} & a. & 0.655$\pm$0.013 & 3480$\pm$14 & 2.501$\pm$0.019 & 2.504$\pm$0.019 \\
 & w. & 0.657$\pm$0.007 & 3479$\pm$8 & 2.502$\pm$0.013 & 2.504$\pm$0.019 \\
\tableline
\multirow{2}{*}{A100} & a. & 0.655$\pm$0.013 & 3479$\pm$14 & 2.501$\pm$0.019 & 2.504$\pm$0.019 \\
 & w. & 0.657$\pm$0.007 & 3479$\pm$8 & 2.503$\pm$0.013 & 2.504$\pm$0.019 \\
\tableline
\multirow{2}{*}{P100} & a. & 0.624$\pm$0.015 & 3543$\pm$14 & 2.456$\pm$0.019 & 2.434$\pm$0.022 \\
 & w. & 0.629$\pm$0.008 & 3543$\pm$8 & 2.456$\pm$0.013 & 2.434$\pm$0.022 \\
\tableline
\multirow{2}{*}{PQS} & a. & 0.630$\pm$0.014 & 3537$\pm$14 & 2.460$\pm$0.018 & 2.443$\pm$0.021 \\
 & w. & 0.635$\pm$0.008 & 3538$\pm$8 & 2.460$\pm$0.013 & 2.443$\pm$0.021 \\
\tableline
all & a. & 0.641$\pm$0.020 & 3510$\pm$32 & 2.480$\pm$0.026 & 2.471$\pm$0.033 \\
\tableline
\end{tabular*}
\begin{tablenotes}
\item The uncertainties in the overall arithmetic means are the standard deviations of the individual parameter values.
\end{tablenotes}
\end{threeparttable}
\end{center}
\end{table*}

\begin{figure}[!ht]
\epsscale{0.99}
\plotone{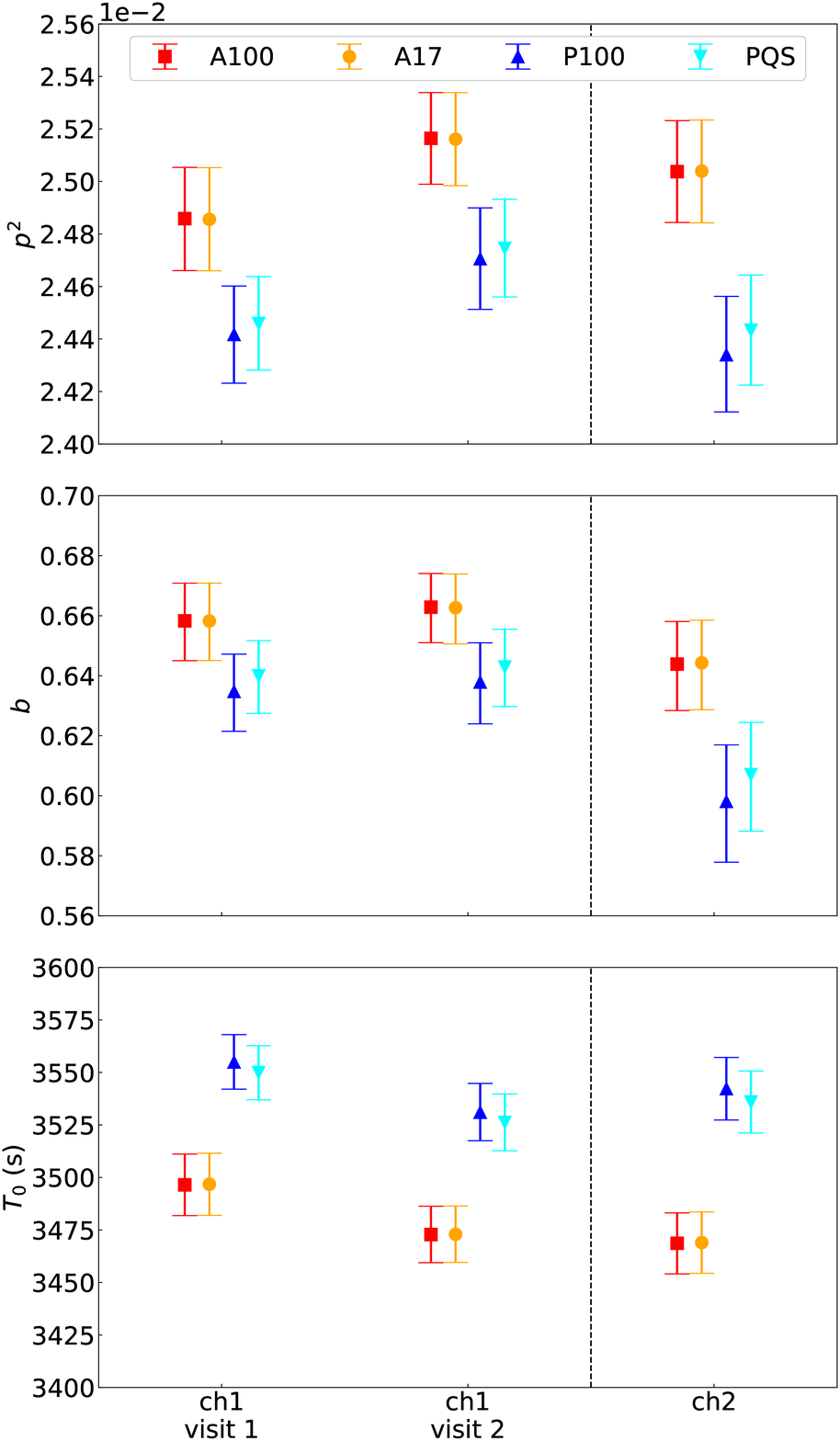}
\caption{Top panel: transit depth estimates obtained with different sets of limb-darkening coefficients: A100 (red squares), A17 (orange circles), P100 (blue, upward triangles), and PQS (cyan, downward triangles). Middle and bottom panels: analogous plots for the impact parameter and for the transit duration.
\label{fig:ld_p2bT0params}}
\end{figure}

\subsection{Transit parameters}
\label{ssec:transit_params}

Figure~\ref{fig:ld_p2bT0params} reports the best-fit transit depth ($p^2$), impact parameter ($b$), and transit duration ($T_0$), obtained with the four sets of limb-darkening coefficients reported in Table~\ref{tab_ldc} (see Section~\ref{ssec:limb-darkening}). There appear to be systematic offsets between the parameters obtained by using the ATLAS and PHOENIX sets of coefficients. In particular, the PHOENIX models lead to smaller transit depths by $\sim$400--700 parts per million (ppm), smaller impact parameters by $\sim$0.02--0.04, and longer transit durations by 55--70 s, at 3.6 and 4.5~$\mu$m, respectively. These differences correspond to two to five times the respective parameter error bars.

We did not find strong evidence in favor of one specific model (see Appendix~\ref{app_ldc}). Table~\ref{tab_means} reports the arithmetic and weighted mean values of the geometric parameters, $b$ and $T_0$, across the three visits for each limb-darkening model, and the mean transit depths at 3.6 and 4.5~$\mu$m. Table~\ref{tab_means} also reports the global mean values over all the different limb-darkening models. While the absolute transit depths are model-dependent, the difference between the 3.6 and 4.5~$\mu$m transit depths is always consistent with zero within 1~$\sigma$.

\section{DISCUSSION}

\subsection{Reliability of the model selection criteria}
\label{ssec:model_selection}
The minimum BIC solution for the second 3.6~$\mu$m visit (quadratic ramp) includes negative nightside flux values, which are unphysical. Even if the minimum nightside flux is consistent with being positive within 1~$\sigma$, the low upper limit poses a challenge for the modeling of exoplanetary atmospheres (e.g., \citealp{kataria15}). The solution obtained by using a linear ramp parameterization, instead of quadratic, appears to be less problematic, as it is discussed in Section~\ref{ssec:EDIBT}.

We tested model selection tools other than the BIC, which all agreed on the choice of the quadratic ramp model, although with different strengths of evidence. In particular, the $\Delta$BIC=8.9 between the linear and quadratic parameterizations (see Table~\ref{tabBICAIC}) denotes a strong, but not conclusive, preference for the latter according to \cite{raftery95}. The Akaike Information Criterion (AIC; \citealp{akaike74}) and the Consistent Akaike Information Criterion (CAIC; \citealp{bozdogan87}) favor the quadratic ramp model more/less strongly than BIC ($\Delta$AIC=15.6, $\Delta$CAIC=7.9), because of a smaller/larger penalty for the number of model parameters. The Deviance Information Criterion \citep{spiegelhalter02} gives results similar to the AIC, while the Bayesian evidences calculated with MultiNest \citep{buchner14} are consistent with the BIC estimates. Therefore, any information criterion weighted average of the alternative models, e.g., the marginalization method proposed by \cite{dewit16}, would be driven by the quadratic ramp. A more sophisticated approach consists of marginalization over the hyperparameters of a Gaussian Process (GP; \citealp{gibson14, evans15}). In this paper, we did not pursue the GP method, because of its high computational cost and unclear performances in previous analyses of the \textit{Spitzer}/IRAC data \citep{ingalls16}.

All of the tests discussed above rely on the relative amplitudes of the residuals. However, it is not guaranteed that smaller residuals correspond to more reliable parameter estimates. The potential errors in the instrumental systematics model may be compensated in part by biasing the retrieved astrophysical parameters, especially if the two sets of parameters (instrumental systematics and astrophysical) are correlated. We observed that, in the second 3.6~$\mu$m data set, the minimum nightside flux is strongly correlated with the two quadratic ramp coefficients, as measured by the absolute value of the Pearson Correlation Coefficients (PCCs $\sim$0.6). The correlation with the linear ramp coefficient is much smaller (PCC$\sim$0.1). 

In conclusion, simple statistical criteria based on the amplitude of the best-fit residuals can provide useful guidelines to model selection, but they should not be considered alone. Physical plausibility may pose important constraints to the model selection, especially when the competing models have similar scores \citep{ingalls16}. We highly recommend to perform some self-consistency tests on the data, e.g., checking that the best-fit parameters do not vary dramatically if analyzing smaller portions rather than the whole data set (see Appendix \ref{app_halfpc}).

\begin{figure*}[!t]
\epsscale{0.95}
\plotone{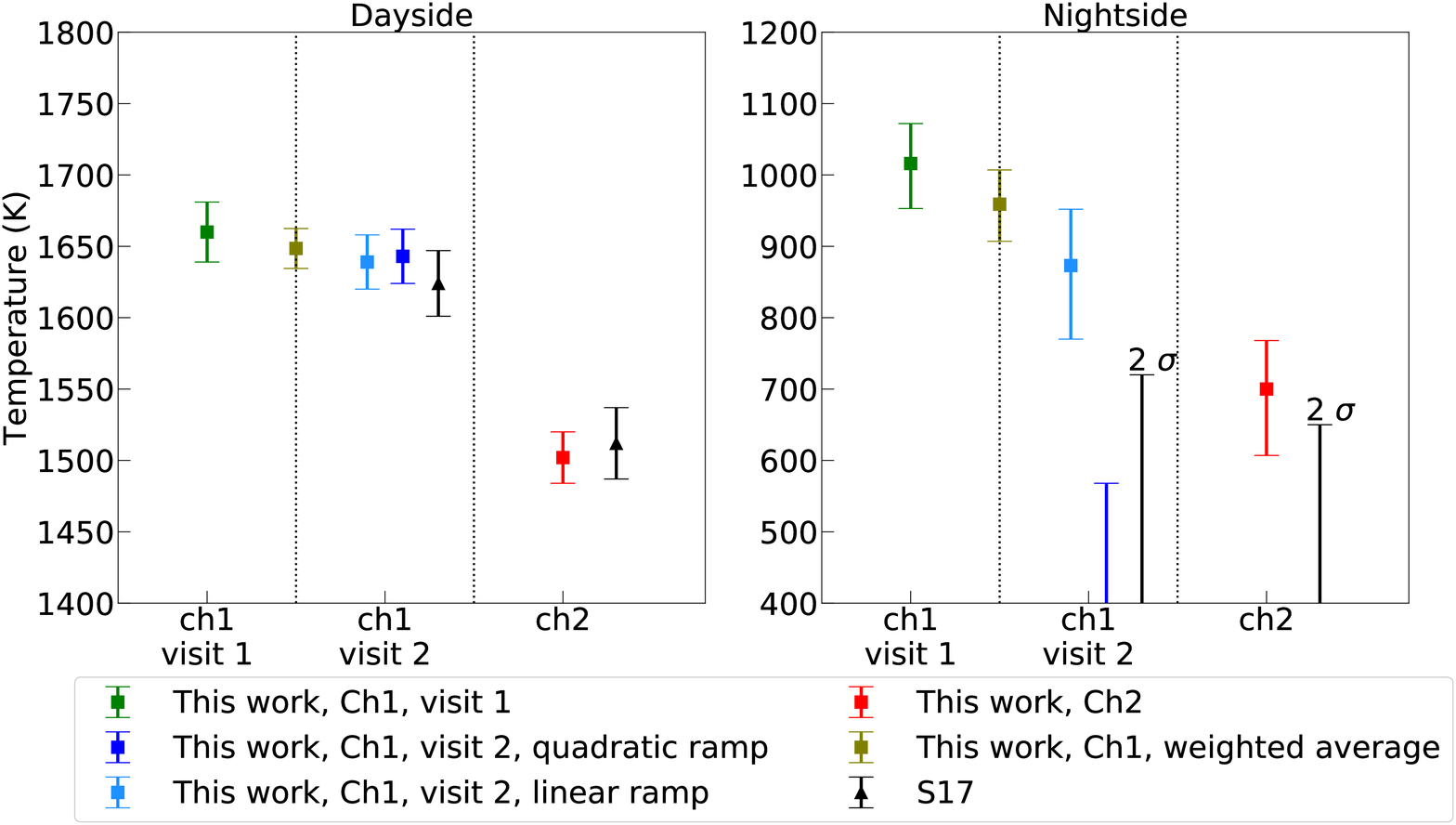}
\caption{Left panel: maximum dayside temperatures obtained in this work for the first 3.6~$\mu$m visit (green), second 3.6~$\mu$m visit using a quadratic (blue) or linear (dodger blue) ramp model, weighted average between the first and second visit with a linear ramp (olive), 4.5~$\mu$m visit (red), and temperatures reported by \citetalias{stevenson17} (black triangles).  Right panel: analogous plot for the minimum nightside temperatures. Note that \citetalias{stevenson17} only reported 2~$\sigma$ upper limits for the nightside temperatures.
\label{fig:brightness_temperatures_compar}}
\end{figure*}

\subsection{Exoplanet Disk-Integrated Brightness Temperatures}
\label{ssec:EDIBT}

For the \textit{Spitzer}/IRAC 3.6 and 4.5~$\mu$m channels, the exoplanet flux contribution is predominantly thermal emission. By neglecting the nonthermal contributions, the observed planet-to-star flux ratio in a given passband is \citep{charbonneau05}
\begin{equation}
\label{eqn:EDIBT}
\frac{F_{p,\lambda }}{F_{*,\lambda }} = \left ( \frac{R_p}{R_*} \right )^2  \frac{ B_{\lambda} (T_p) }{ B_{\lambda} (T_*) } ,
\end{equation}
where $(R_p/R_*)^2$ is the sky-projected planet-to-star area ratio, $T_p$ is the phase-dependent exoplanet brightness temperature, and $T_*$ is the star brightness temperature. We computed the brightness temperatures for a star with $T_{\mbox{\footnotesize eff}} =$4400~K, $\log{g} =$4.65 (see Table~\ref{tab1}), in the two IRAC passbands, by interpolating on a grid of PHOENIX stellar-atmosphere models \citep{husser13}. By inverting Equation~\ref{eqn:EDIBT}, we calculated the exoplanet maximum dayside temperature, $T_{\mbox{\footnotesize day}}^{\mbox{\footnotesize MAX}}$, to be 1660$\pm$21~K for the first 3.6~$\mu$m visit, 1643$\pm$19~K (quadratic ramp) or 1639$\pm$19~K (linear ramp) for the second 3.6~$\mu$m visit, and 1502$\pm$18~K at 4.5~$\mu$m. The corresponding minimum nightside temperatures, $T_{\mbox{\footnotesize night}}^{\mbox{\footnotesize MIN}}$, are 1016$_{-\mbox{63}}^{+\mbox{56}}$~K, $<$568~K, 837$_{-\mbox{103}}^{+\mbox{79}}$~K, and 700$_{-\mbox{93}}^{+\mbox{68}}$~K. These temperature estimates can be visualized in Figure~\ref{fig:brightness_temperatures_compar}.

The choice of a linear or a quadratic ramp parameterization for the second 3.6~$\mu$m data set changes dramatically the inferred astrophysical scenario. The former leads to consistent measurements between the two 3.6~$\mu$m visits  within 1.5~$\sigma$, and lower brightness temperatures at 4.5~$\mu$m. The lower brightness temperatures suggest higher absorption/scatter at 4.5~$\mu$m within the WASP43~b atmosphere, assuming non-inverted thermal profile \citep{blecic14}.

The quadratic ramp model implies lower nightside flux, and brightness temperature (upper limit), for the second 3.6~$\mu$m visit, suggesting some variability with 2.5~$\sigma$ significance level. The wavelength--temperature trend is inverted between the exoplanet dayside and nightside, which indicates different atmospheric opacities between the two sides. These results might be explained with the appearance of high-altitude clouds in the exoplanet atmosphere during the second 3.6~$\mu$m visit. In conclusion, we cannot rule out this solution as physically impossible, but it is most likely biased by the strong parameter correlations, as mentioned in Section~\ref{ssec:model_selection}. This idea is reinforced by the simpler physical interpretation (and smaller parameter correlations) associated with the alternative solution using a linear ramp model.

\begin{figure*}[!t]
\epsscale{0.95}
\plotone{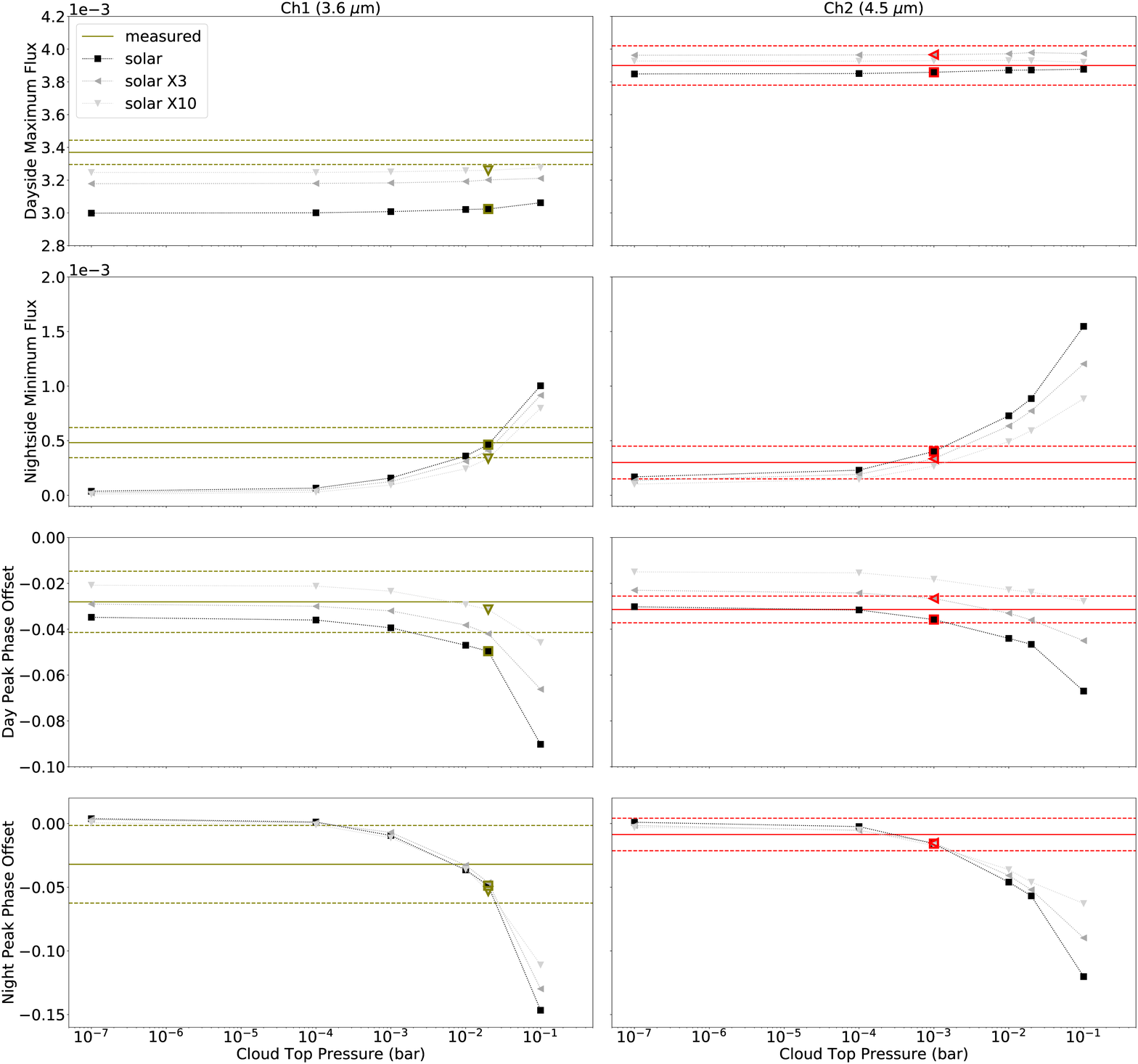}
\caption{Measured phase curve parameters as reported in Table~\ref{tab_pc_final}: the best parameter  values are represented by continuous horizontal lines, the limits of the 1~$\sigma$ interval are represented by dashed horizontal lines. Phase curve parameters obtained from the atmospheric models with 1$\times$ (black squares), 3$\times$ (dark gray left-pointing triangles), and 10$\times$ (light gray downward-pointing triangles) solar metallicity as a function of the cloud top pressure (10$^{-1}$ bar corresponds to the cloud-free models). The highlighted points correspond to the best matching phase curve models represented in Figure~\ref{fig:pc_atmo_curves}.
\label{fig:pc_atmo_params}}
\end{figure*}

\begin{figure*}[!ht]
\epsscale{0.95}
\plotone{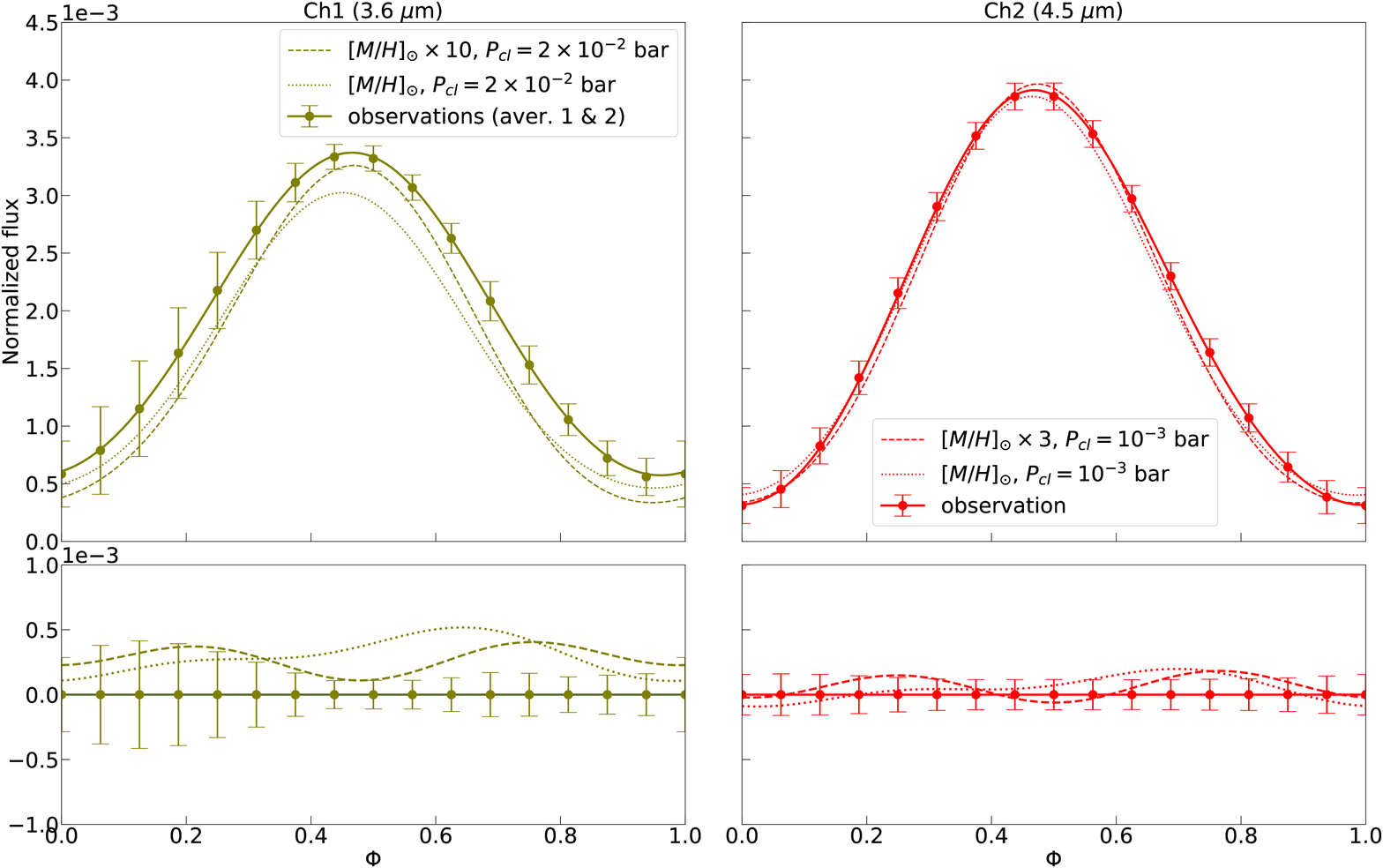}
\caption{Left, top panel: observed 3.6~$\mu$m phase curve profile (continuous line and points with error bars), i.e., average of the best-fit profiles for the two visits (corresponding to the green and dodger blue curves in Figure~\ref{fig:phasecurves_best}), the best match from our grid of atmospheric phase curve models (dashed line), and the best match with solar metallicity (dotted line). The larger error bars for $\Phi <$0.5 take into account the discrepancy between the best-fit profiles for the two visits. Left, bottom panel: residuals between the observed and two model phase curves. Right panels: analogous plots for the 4.5~$\mu$m phase curve.
\label{fig:pc_atmo_curves}}
\end{figure*}

\subsection{Atmospheric circulation models}
\label{ssec:pc_atmo}
We used the \texttt{2D-ATMO} code \citep{tremblin17} to compute a grid of phase curve models for WASP43~b. The \texttt{2D-ATMO} is an extension of \texttt{1D-ATMO} \citep{tremblin15} that takes into account the circulation induced by the irradiation from the host star at the equator of the planet. The atmospheric model for WASP43~b has been computed as part of a model comparison performed for future JWST observations (Venot et al., in preparation).  The magnitude of the zonal wind is imposed at the substellar point at 4~km~s$^{-1}$ and is computed accordingly to the momentum conservation law in the rest of the equatorial plane. The vertical mass flux is assumed to be proportional to the meridional mass flux with a proportionality constant $1/\alpha$; the wind is therefore purely longitudinal and meridional if $\alpha \rightarrow \infty$ or purely longitudinal and vertical for $\alpha \rightarrow 0$. 
As in \cite{tremblin17}, a relatively low value of $\alpha$ drives the vertical advection of entropy/potential temperature in the deep atmosphere that can produce a hot interior, which can explain the inflated radii of hot Jupiters. A high value of $\alpha$ will produce a cold deep interior as in the standard 1D models. In this study, we used a simulation with $\alpha$=10$^4$ that should be representative of WASP43~b since the planet is not highly inflated. In order to reproduce the low fluxes on the nightside of the planet, we added a simple cloud model consisting of a gray absorbing cloud deck with an absorption of 2.5~m$^2$~kg$^{-1}$ with a fixed bottom pressure of 0.1 bar. We explored different metallicities (1$\times$, 3$\times$, and 10$\times$ solar) and different top pressure levels for the cloud deck (0.1 bar, i.e., no clouds, 0.02, 0.01, 10$^{-3}$, 10$^{-4}$ and 10$^{-7}$ bar).

Figure~\ref{fig:pc_atmo_params} compares the measured day and nightside fluxes and peak offsets (from Table~\ref{tab_pc_final}) with those predicted by the atmospheric models. Figure~\ref{fig:pc_atmo_curves} compares the whole phase curve profiles obtained from the data with the best matching profiles from the atmospheric models.

A number of atmospheric phase curve models are in excellent agreement with the observed profile at 4.5~$\mu$m. The best matches are the models with 3$\times$ or 1$\times$ solar metallicity and cloud top pressure of 10$^{-3}$ bar. In both cases, the discrepancies between the fitted and the model profiles are smaller than 200~ppm with rms amplitudes of $\sim$100~ppm. The corresponding model phase curve parameters are all consistent with the measured values within 1~$\sigma$. In general, all of the models with cloud top pressure lower than 10$^{-2}$ bar are in good agreement with the 4.5~$\mu$m observation, but the models with 10$\times$ solar metallicity tend to predict a smaller dayside peak offset.
The models with no clouds can be ruled out at the 4-8~$\sigma$ level in the nightside flux, and they also tend to predict significantly larger peak offsets, depending on the metallicity.

The results at 3.6~$\mu$m are more problematic, as the measured dayside flux is higher than predicted by the atmospheric models. The best match to the observed profile (average of the two observations) is the model with 10$\times$ solar metallicity and cloud top pressure of 2$\times$10$^{-2}$ bar. In this case, the discrepancies between the fitted and the model profiles are within $\sim$400~ppm with rms amplitudes below 300~ppm, which are larger than the error bars of the best-fit profile at certain orbital phases. The corresponding model phase curve parameters are consistent with the measured values within 1.5~$\sigma$. The models with 3$\times$ and 1$\times$ solar metallicity predict smaller than measured dayside fluxes by $\sim$150~ppm (2~$\sigma$) and 300~ppm (4~$\sigma$), respectively.

The observations at 3.6 and 4.5~$\mu$m are best described by atmospheric models with different metallicity and cloud top pressure, although a range of models with metallicity higher than solar and cloud top pressure of $\sim$10$^{-2}$ bar can reproduce all of the measured phase curve parameters within less than 2~$\sigma$. It is likely that a chemical composition different than scaled solar abundances could provide a better match to the data, without the need to find a compromise at the edges of the acceptable parameter ranges for the observations at the two wavelengths. In this paper, we refrain from speculating about the possible nature of the non-standard chemistry in the atmosphere of WASP43~b, which cannot be probed with the current data.

As the observations were not taken simultaneously, we cannot exclude some variability of the nightside clouds over the different visits. In Section~\ref{ssec:pc_results} we noted that the parameter results derived from the individual 3.6~$\mu$m visits were not fully consistent at the 1~$\sigma$ level, but the apparent discrepancies might be caused by correlated noise in the fitting residuals. In general, the observations with 4.5~$\mu$m channel are much less affected by correlated noise \citep{krick16}, which is also confirmed by the analyses in this paper (see Section~\ref{sec:results} and Appendix~\ref{app_halfpc}). Therefore, new observations at 4.5~$\mu$m would help to assess the  level of variability in the atmosphere of WASP43~b.

\begin{figure}[!t]
\epsscale{0.99}
\plotone{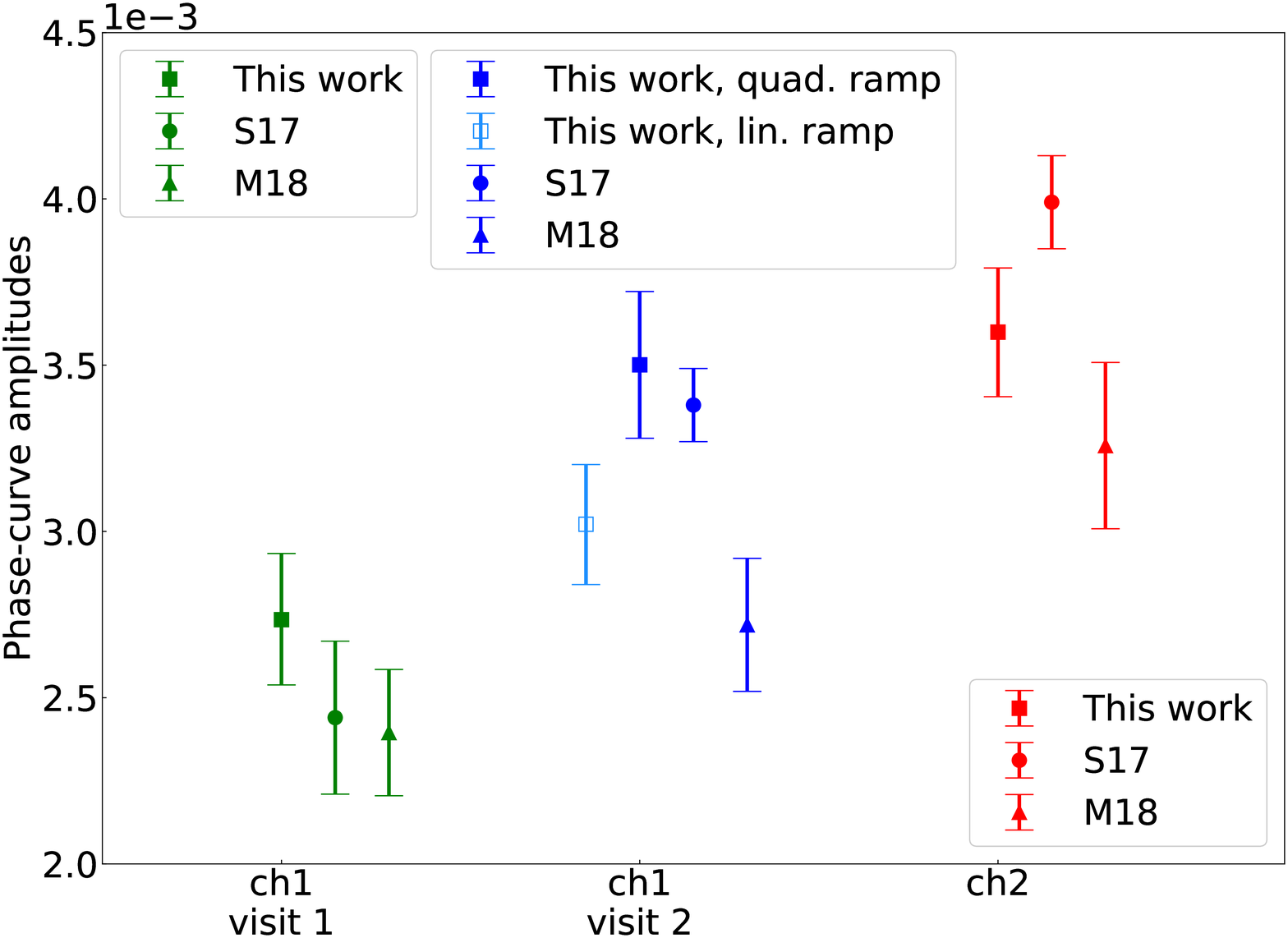}
\caption{Peak-to-peak phase curve amplitudes obtained in this work (squares), and reported by \citetalias{stevenson17} (circles) and \citetalias{mendonca18} (triangles). Same choice of colors as in Figures~\ref{fig:daynight_best_withcomparstev} and \ref{fig:brightness_temperatures_compar}.
\label{fig:peak_to_peak_amplitudes_compar}}
\end{figure}

\subsection{Comparison with previous analyses of the same observations}
The rms amplitudes of the light curve fitting residuals obtained with our wavelet pixel-ICA are within 1$\%$ of those reported by \citetalias{mendonca18}, using an extension of the BLISS mapping algorithm \citep{stevenson12}.

Figure~\ref{fig:peak_to_peak_amplitudes_compar} reports the phase curve peak-to-peak amplitudes that we computed to compare with those reported in the previous literature. When taking the linear ramp model for the second 3.6~$\mu$m visit, our best-fit amplitudes are consistent with those reported by \citetalias{mendonca18} within 1~$\sigma$, though our central values are $\sim$300~ppm larger. \citetalias{stevenson17} obtained larger peak-to-peak amplitudes for the second 3.6~$\mu$m and 4.5~$\mu$m visits. When taking the quadratic ramp model for the second 3.6~$\mu$m visit, we also obtain a larger peak-to-peak amplitude, in agreement with \citetalias{stevenson17}. Overall, the different estimates for each visit are consistent at the 2~$\sigma$ level.

We observed that, in the analyses discussed here, a smaller peak-to-peak amplitude corresponds to a higher nightside flux (at a given wavelength). Figure~\ref{fig:brightness_temperatures_compar} compares the brightness temperatures obtained in this work with those reported by \citetalias{stevenson17}. The dayside temperatures are consistent within 0.5~$\sigma$. Our estimates of the nightside temperatures are higher than the 2~$\sigma$ upper limits reported by \citetalias{stevenson17}. 
Figure~\ref{fig:daynight_best_withcomparstev} shows that we obtained a significantly smaller dayside peak offset compared to \citetalias{stevenson17} at 4.5~$\mu$m.

\citetalias{stevenson17} could not find adequate approximations to the observed phase curve profiles with the cloud-free atmospheric models of \cite{kataria15}. \citetalias{mendonca18} computed new global circulation models with \texttt{THOR} \citep{mendonca16}; their best match to the \textit{Spitzer}/IRAC data is a model with a nightside cloud deck with top pressure of 10$^{-2}$~bar and enhanced carbon dioxide (CO$_2$) with a longitudinal gradient. A visual inspection of the Figure~6 in \citetalias{mendonca18} reveals that the maximum discrepancies between the fitted and the best model phase curves are $\sim$400--500~ppm, i.e., equal or larger than those obtained in this study (see Section~\ref{ssec:pc_atmo}). Furthermore, the non-equilibrium CO$_2$ was introduced ad hoc by \citetalias{mendonca18} to reproduce the low nightside flux observed at 4.5~$\mu$m, with a lower limit for the cloud top pressure of 10$^{-2}$~bar (from \citealp{kreidberg14}). \cite{mendonca18b} could not physically explain that chemical disequilibrium.

\begin{figure*}[!t]
\epsscale{0.95}
\plotone{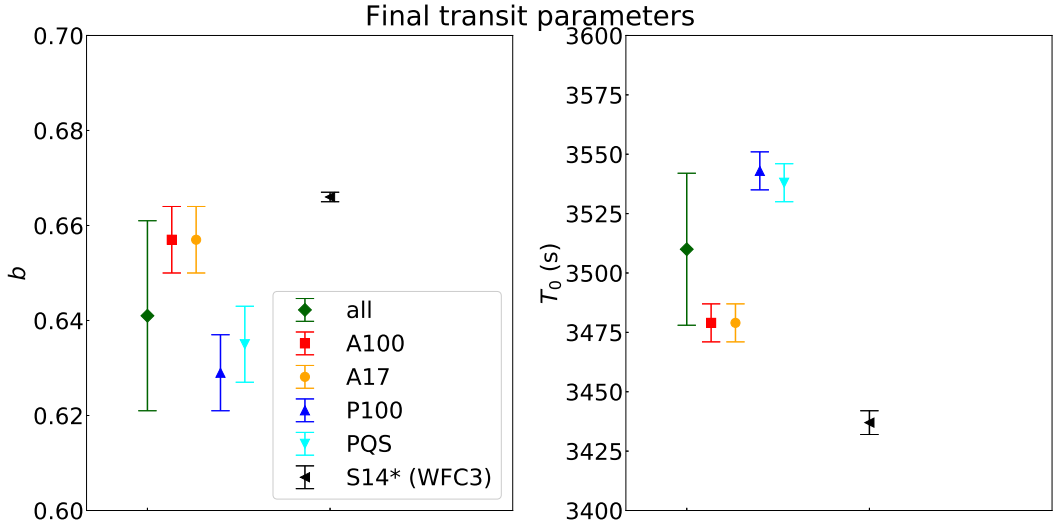}
\caption{Left panel: impact parameter estimates obtained in this work (Table~\ref{tab_means}) and in the previous literature. This work: weighted mean values over the three observations by using A100 (red square), A17 (orange circle), P100 (blue, upward triangle), and PQS (cyan, downward triangle) limb-darkening coefficients, and overall arithmetic mean (green diamond). The S14 result (black, leftward triangle) has been calculated, in this work, from the other parameters reported by \cite{stevenson14}. Right panel: analogous plot for the transit duration.
\label{fig:bT0_literature}}
\end{figure*}

Figure~\ref{fig:bT0_literature} reports our final estimates of the geometric parameters averaged over the three observations, and the analogous parameters derived from those reported by \cite{stevenson14} using \textit{HST}/WFC3 data.
Our estimates of the impact parameter by using ATLAS limb-darkening coefficients are marginally consistent within 1~$\sigma$ with the value reported by \cite{stevenson14}. We found longer transit duration than \cite{stevenson14} by 42~s and 101--106~s by using ATLAS and PHOENIX limb-darkening coefficients, respectively. These discrepancies are above 3~$\sigma$. Note that the small uncertainties in the parameters obtained from \cite{stevenson14} do not account for the degeneracy with the stellar limb-darkening.

\begin{figure*}[!ht]
\epsscale{0.95}
\plotone{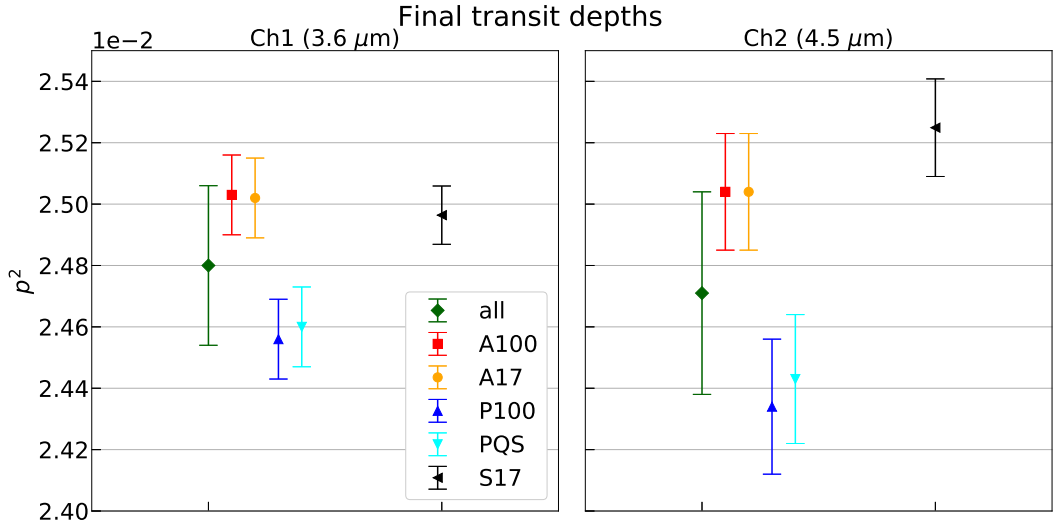}
\caption{3.6 and 4.5~$\mu$m transit depths estimates obtained in this work (Table~\ref{tab_means}) and by \citetalias{stevenson17}. Same choice of colors and symbols as in Figure~\ref{fig:bT0_literature}.
\label{fig:p2_literature}}
\end{figure*}

Figure~\ref{fig:p2_literature} compares the 3.6 and 4.5~$\mu$m transit depths with those obtained by \citetalias{stevenson17}. Unsurprisingly, our estimates using ATLAS (claret-4) limb-darkening coefficients best match the results obtained by \citetalias{stevenson17}, which also adopted ATLAS (quadratic) limb-darkening coefficients.


\subsection{Comparison with other observations and with other exoplanets}

Some authors suggested the existence of a simple relationship between the irradiation temperature and circulation efficiency of the exoplanetary atmospheres \citep{cowan11, perez-becker13, schwartz15, komacek16, keating17}. The \citetalias{stevenson17} claim of zero circulation efficiency for the WASP43~b atmosphere injected an apparent outlier to the expected trend. WASP43~b and HD209458~b have similar irradiation temperatures of $\sim$2000--2100~K. \cite{schwartz17} reported $\varepsilon =$0.49$_{-0.14}^{0.15}$ for HD209458~b, based on visible-to-infrared observations, often limited to the secondary eclipses.

In this work, we obtained significantly higher nightside temperatures than the previous estimates by \citetalias{stevenson17} for WASP43~b in the \textit{Spitzer}/IRAC passbands. Assuming a blackbody-like emission, the circulation efficiency goes up to $\varepsilon \sim$0.1--0.3. We emphasize that this is just a broad estimate of the circulation efficiency in the WASP43~b atmosphere. In fact, the blackbody assumption is not valid, as revealed by the $>$4.5~$\sigma$ difference between the 3.6 and 4.5~$\mu$m dayside temperatures (see Section~\ref{ssec:EDIBT}). We propose a more direct comparison between the 4.5~$\mu$m phase curves of WASP43~b and HD209458~b. \cite{zellem14} reported $T_{\mbox{\footnotesize day}} =$1499$\pm$15~K and $T_{\mbox{\footnotesize night}} =$972$\pm$44~K for HD209458~b at 4.5~$\mu$m. We obtained the same dayside temperature (within 0.2~$\sigma$) and $\sim$200--300~K lower nightside temperature for WASP43~b at the same wavelength. These comparisons suggest the WASP43~b atmosphere may have a lower circulation efficiency than HD209458~b, but the difference appears to be significantly smaller than from the original estimates reported in the literature. Furthermore, there are some hints of variability in the nightside cloud deck of WASP43~b (see Section~\ref{ssec:pc_atmo}), that would affect the temperature measurements. A new set of observations is desirable to test this hypothesis.

\begin{figure*}[!ht]
\epsscale{1.0}
\plotone{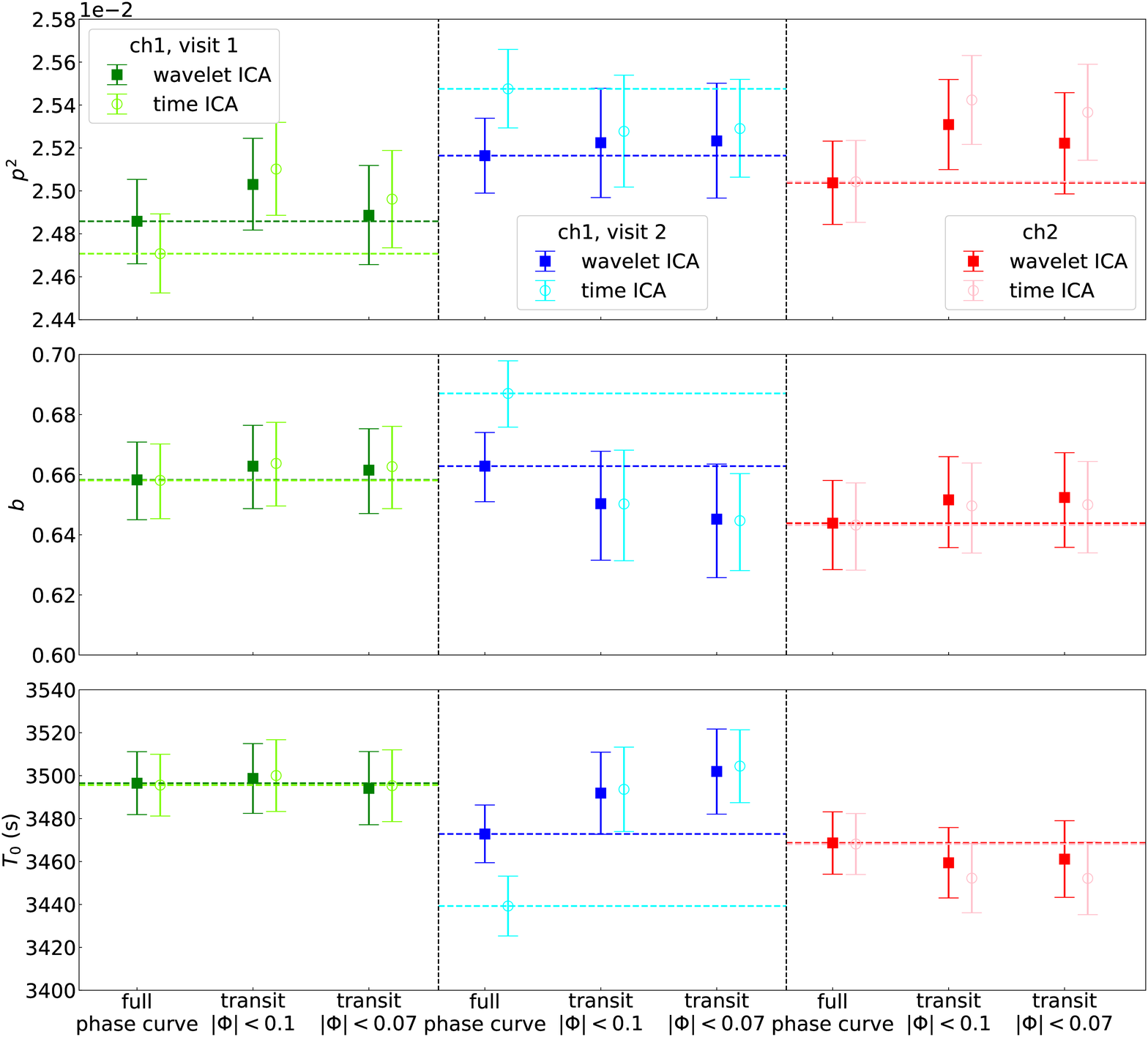}
\caption{Top panel: transit depth estimates obtained from the full phase curve, and from two transit-only analyses with different baselines (see Section~\ref{sec:transit}) by using the wavelet pixel-ICA (darker colors) and time pixel-ICA techniques (lighter colors). The horizontal dashed lines correspond to the full phase curve results. Middle and bottom panels: analogous plots for the impact parameter and transit duration.
\label{fig:p2bT0_phase_tran_A100}}
\end{figure*}

\section{Transit-only analyses}
\label{sec:transit}
We analyzed smaller portions of the data sets as transit-only observations, in order to evaluate pros and cons of the different observation types for exoplanet characterization. We tested two phase intervals, $| \Phi | \le 0.1$ and $| \Phi | \le 0.07$, which correspond to $\sim$3.3 and 2.3 times the full transit duration. The fitted transit model does not include any phase curve modulation or exoplanet nightside pollution, as is common practice in transit-only observations.

Figures~\ref{fig:p2bT0_phase_tran_A100} compares the transit parameters obtained from the phase curve and transit-only analyses. We found that the parameter error bars scale approximately as $\sqrt{N_{\mbox{\footnotesize tot}}/N_{\mbox{\footnotesize out}}}$, where $N_{\mbox{\footnotesize tot}}$ is the total number of data points, and $N_{\mbox{\footnotesize out}}$ is the number of out-of-transit points. The mathematical derivation of this result is reported in Appendix~\ref{app_scaling_error} (Equation~\ref{eqn:transit_depth_error_approx}).
More specifically, the error bars in transit depth obtained for the first 3.6 and 4.5~$\mu$m visits with the transit-only analyses are 9$\%$ and 8$\%$ (longer configuration), and 17$\%$ and 22$\%$ (shorter configuration) larger than those obtained from the respective full phase curves, which are within the ranges predicted by Equation~\ref{eqn:transit_depth_error_approx} (8--16$\%$ and 13--29$\%$, see Appendix~\ref{app_scaling_error}). For the second 3.6~$\mu$m observation, the transit-only error bars are larger than expected, i.e., 46$\%$ and 53$\%$ larger than the respective full phase curve error bars. It is worth noting that the second 3.6~$\mu$m observation is the only one that required a non-constant ramp model in the phase curve analysis. The error bars (in transit depth) obtained with the shorter transit-only configuration are 6--12$\%$ larger than those obtained with the longer transit-only configuration for all of the observations, in good agreement with the range predicted by Equation~\ref{eqn:transit_depth_error_approx} (4--10$\%$, see Appendix~\ref{app_scaling_error}). The other transit parameters have similar differences between the error bars obtained with the various configurations.

Interestingly, the transit depth estimated from the (wavelet) transit-only analyses, and especially from those with the longer phase interval, are slightly larger than those obtained from the full phase curve analyses in all cases. Although the three transit depth estimates for the same observation are mutually consistent within 1~$\sigma$, we found that the observed systematic behavior can be caused by the flat baseline approximation. In fact, the phase-dependent exoplanetary flux is higher before and after transit than during the transit, therefore increasing the apparent transit depth if this effect is not taken into account. Based on the best-fit phase curve model, the differences between the time-averaged exoplanetary flux out-of-transit and in-transit are in the range of 40--120~ppm. The differences in transit depths are of the same order of magnitude, but not identical because of the correlations with the other free parameters in the fit, which are also slightly biased. Consequently, the largest differences in transit depth are obtained for the 4.5~$\mu$m visit, i.e., +271 and +185~ppm for the longer and shorter transit-only analyses, respectively. The parameter offsets decrease significantly if the best-fit phase curve parameters are fixed in the transit-only analyses.
While these potential bias are not statistically significant with the current error bars, they might become significant with the smaller error bars that are expected to be achieved with the next-generation instruments, such as those onboard JWST and ARIEL. The potential bias are expected to be smaller for exoplanets with larger orbits, both because of the smaller day-night temperature contrast (phase curve amplitude) and the longer orbital period relative to the transit duration.

\begin{figure*}[!t]
\epsscale{0.95}
\plotone{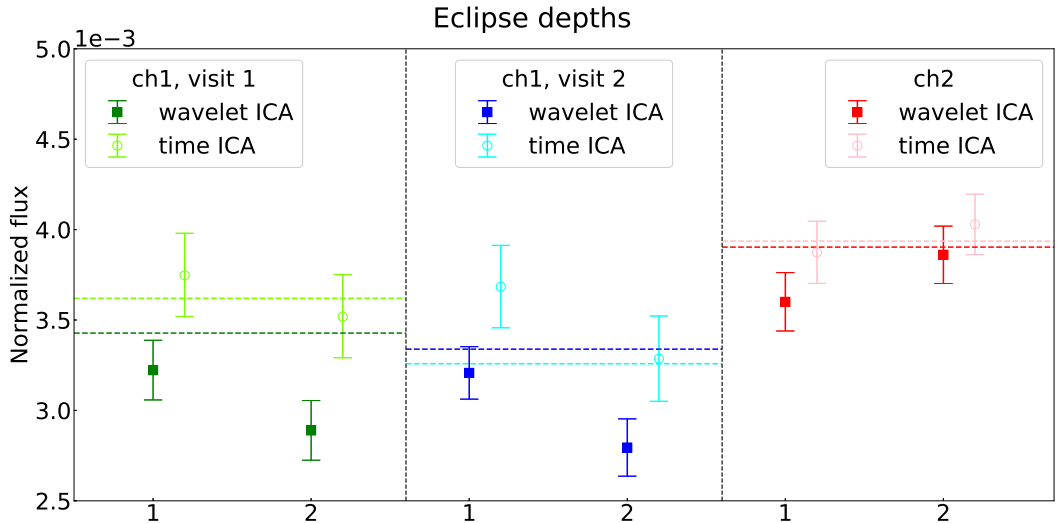}
\caption{Eclipse depth estimates from the eclipse-only analyses by using the wavelet pixel-ICA (darker colors) and time pixel-ICA techniques (lighter colors). The horizontal dashed lines act as upper limits, i.e., the flux maxima obtained from the full phase curve analyses. The coordinates 1 and 2 refer to the first and second eclipse of each visit.
\label{fig:eclipse_depths_philim0e1}}
\end{figure*}

\section{Eclipse-only analyses}
\label{sec:eclipse}
We also analyzed portion of the data sets as eclipse-only observations. We fixed the orbital parameters to the estimates obtained from the corresponding transit-only observation, then fitted for the eclipse depth and timing. There are no analogous eclipse depth estimates for the full phase curve analyses, as the out-of-eclipse flux is not constant, but the dayside maxima should represent upper limits for the eclipse depths. Figure~\ref{fig:eclipse_depths_philim0e1} reports the eclipse depth estimates for the phase interval  $| \Phi -0.5 |\le 0.1$, and the dayside maxima from the full phase curve analyses. The eclipse depths obtained with wavelet pixel-ICA at the same wavelength are mutually consistent within 1.5~$\sigma$, and they are all below the respective phase curve upper limits.

\section{CONCLUSIONS}
We analyzed three \textit{Spitzer}/IRAC phase curves of the exoplanet WASP43~b at 3.6~$\mu$m (two observations) and 4.5~$\mu$m using a blind signal-source separation method, i.e., the wavelet pixel-ICA. We assessed the robustness of the results by analysing both the full and the half phase curves with different instrument ramp models. We revealed a significant degeneracy between stellar limb-darkening and transit parameters, and a potential bias obtained analyzing only the transit portion of the phase curve. This bias is of the order of $\sim$100~ppm in transit depth for WASP43~b in the mid-infrared, and it is expected to be highly significant for the upcoming JWST and ARIEL observations. We found higher nightside temperatures, smaller hotspot offsets, and greater consistency ($\sim$1~$\sigma$) between the two 3.6~$\mu$m visits than those reported by \cite{stevenson17}. Our results point toward a greater circulation efficiency of the WASP43~b atmosphere, in agreement with an empirical trend between irradiation temperature and circulation efficiency. Additionally, we compared the observed phase curves with a grid of atmospheric models, enabling quantitative estimates of the cloud top pressure. Our phase curve parameter results are consistent within 1~$\sigma$ with those reported in a recent reanalysis by \cite{mendonca18}, but we provide an alternative interpretation with a lower cloud top pressure instead of invoking a strong disequilibrium chemistry.
Furthermore, we proposed a simple formula for estimating how the error bars scale with the duration of the observations. Such formula can be used for optimizing the trade-off between parameters precision and duration of the observations.

\acknowledgments
This work was supported by the LabEx P2IO and the French ANR contract 05-BLAN-NT09-573739. The research leading to these results has received funding from the European Union’s Horizon 2020 Research and Innovation Programme, under Grant Agreement n° 776403.
This research has made use of the NASA/IPAC Infrared Science Archive, which is operated by the Jet Propulsion Laboratory, California Institute of Technology, under contract with the National Aeronautics and Space Administration.

\clearpage

\appendix
\section{Scaling relation for the error bars in transit depth}
\label{app_scaling_error}
We derive here a simple analytical formula to estimate the error bar relative to the
transit depth as a function of the time spent observing the out-of-transit. We consider a simplified case, with flat out-of-transit and in-transit, no stellar limb-darkening, and neglecting the transit ingress and egress. In this case, the transit depth is equal to
\begin{equation}
\label{eqn:transit_depth_simple}
p^2 = \frac{F_{\mbox{\footnotesize out}}-F_{\mbox{\footnotesize in}}}{F_{\mbox{\footnotesize out}}} = 1 - \frac{F_{\mbox{\footnotesize in}}}{F_{\mbox{\footnotesize out}}},
\end{equation}
where $F_{\mbox{\footnotesize in}}$ and $F_{\mbox{\footnotesize out}}$ are the constant flux values in-transit and out-of-transit, respectively. If the data are only affected by gaussian noise, the flux values can be estimated with the following error bars:
\begin{equation}
\label{eqn:dFin_dFout}
\Delta F_{\mbox{\footnotesize in}} = \frac{\sigma }{\sqrt{N_{\mbox{\footnotesize in}}}}, \ \Delta F_{\mbox{\footnotesize out}} = \frac{\sigma }{\sqrt{N_{\mbox{\footnotesize out}}}},
\end{equation}
where $\sigma$ is the standard deviation of the gaussian noise, $N_{\mbox{\footnotesize in}}$ and $N_{\mbox{\footnotesize out}}$ are the numbers of in-transit and out-of-transit data points, respectively.

We calculate the error bar in transit depth, $p^2$, by using the ``law of propagation of error'' \citep{taylor96}:
\begin{eqnarray}
\label{eqn:transit_depth_error_formal}
\nonumber
\Delta p^2 = \sqrt{ \left ( \frac{\partial p^2}{\partial F_{\mbox{\footnotesize in}}} \Delta F_{\mbox{\footnotesize in}} \right )^2 + \left ( \frac{\partial p^2}{\partial F_{\mbox{\footnotesize out}}} \Delta F_{\mbox{\footnotesize out}} \right )^2} = \\
= \sqrt{ \left ( - \frac{\Delta F_{\mbox{\footnotesize in}}}{F_{\mbox{\footnotesize out}}}  \right )^2 + \left ( \frac{F_{\mbox{\footnotesize in}}}{F_{\mbox{\footnotesize out}}^2} \Delta F_{\mbox{\footnotesize out}} \right )^2}.
\end{eqnarray}
By injecting Equation~\ref{eqn:dFin_dFout} into Equation~\ref{eqn:transit_depth_error_formal}, we obtain
\begin{equation}
\label{eqn:transit_depth_error_exact}
\Delta p^2 = \frac{\sigma}{ F_{\mbox{\footnotesize out}}} \sqrt{ \frac{1}{ N_{\mbox{\footnotesize in}}} + \frac{ F_{\mbox{\footnotesize in}}^2 }{  F_{\mbox{\footnotesize out}}^2 } \frac{1}{ N_{\mbox{\footnotesize out}}} }.
\end{equation}
Now, we make the approximation $F_{\mbox{\footnotesize in}} \approx F_{\mbox{\footnotesize out}} = F$, obtaining
\begin{equation}
\label{eqn:transit_depth_error_approx}
\Delta p^2 \approx \frac{\sigma}{F} \sqrt{ \frac{1}{ N_{\mbox{\footnotesize in}}} + \frac{1}{ N_{\mbox{\footnotesize out}}} } = \frac{\sigma}{F} \sqrt{ \frac{ N_{\mbox{\footnotesize tot}} }{ N_{\mbox{\footnotesize in}} N_{\mbox{\footnotesize out}} } }.
\end{equation}
We estimate that, for typical transit depth values up to $\sim$3$\%$, the impact of this approximation is less than 0.1$\%$ in $\Delta p^2$.

The formula in Equation~\ref{eqn:transit_depth_error_approx} should provide a lower limit for the error bars. In a more realistic case, the error bars will be larger because of the non-flatness introduced by the stellar limb-darkening and the phase curve modulations, and, in general, because of correlations between the free parameters in the fit. In this work, we found that the error bars in transit depth are $\sim$20--50$\%$ larger than those estimated using Equation~\ref{eqn:transit_depth_error_approx} with $N_{\mbox{\footnotesize in}}$ ranging from the number of data points between the second to third contact points and the number of data points between the first to fourth contact points \citep{seager03}.

Equation~\ref{eqn:transit_depth_error_approx} provides a useful tool to predict how the error bars can scale with the longer observations, then to optimize the trade-off between observing time and precision with the future missions.

\section{Time vs wavelet pixel-ICA}
\label{app_wavetime}
The core of the pixel-ICA method is the ICA transform of a set of pixel time series into maximally independent components, i.e., a linear transformation that minimizes the mutual information \citep{hyvarinen00}. In the wavelet pixel-ICA algorithm the pixel time series undergo Discrete Wavelet Transform (DWT) before the ICA separation, and the independent components are transformed back into the time domain. More specifically, we adopt one-level DWT with mother wavelet Daubechies-4 \citep{daubechies92}.

One of the independent components has the morphology of the astrophysical signal (transit, eclipse, or phase curve), the other components represent the instrumental systematics. We model fit the sum-of-pixel time series, the so-called raw light curve, as a linear combination of a parametric model of the astrophysical signal (instead of the relevant independent component) and the other independent components.

The MCMC error bars are then rescaled as
\begin{equation}
\sigma_{par} = \sigma_{par,0} \sqrt{ \frac{\sigma_0^2+\sigma_{ICA}^2}{\sigma_0^2} } ,
\end{equation}
where $\sigma_0^2$ is the likelihood variance, approximately equal to the variance of the residuals, and $\sigma_{ICA}^2$ is a term accounting for the uncertainty in the ICA components.
The latter term is calculated as \citep{morello16}
\begin{equation}
\label{eqn:sigmaica}
\sigma_{ICA}^{2} = \sum_j o_j^2 ISR_j ,
\end{equation}
where ISR is the so-called Interference-to-Signal-Ratio matrix \citep{tichavsky08}, and $o_j$ are the best-fit coefficients of the linear combination.

\begin{figure*}[!h]
\epsscale{0.95}
\plotone{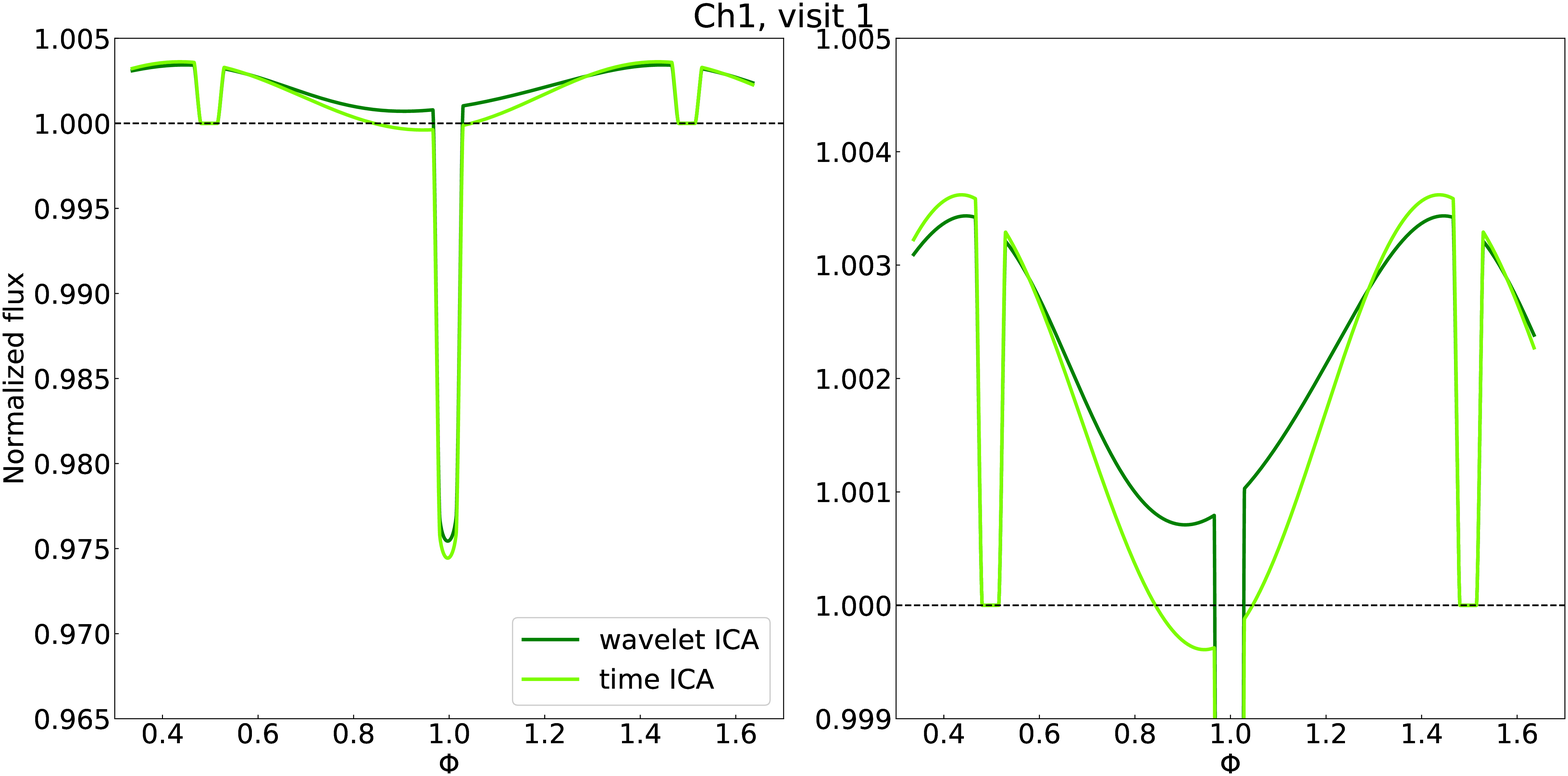}
\plotone{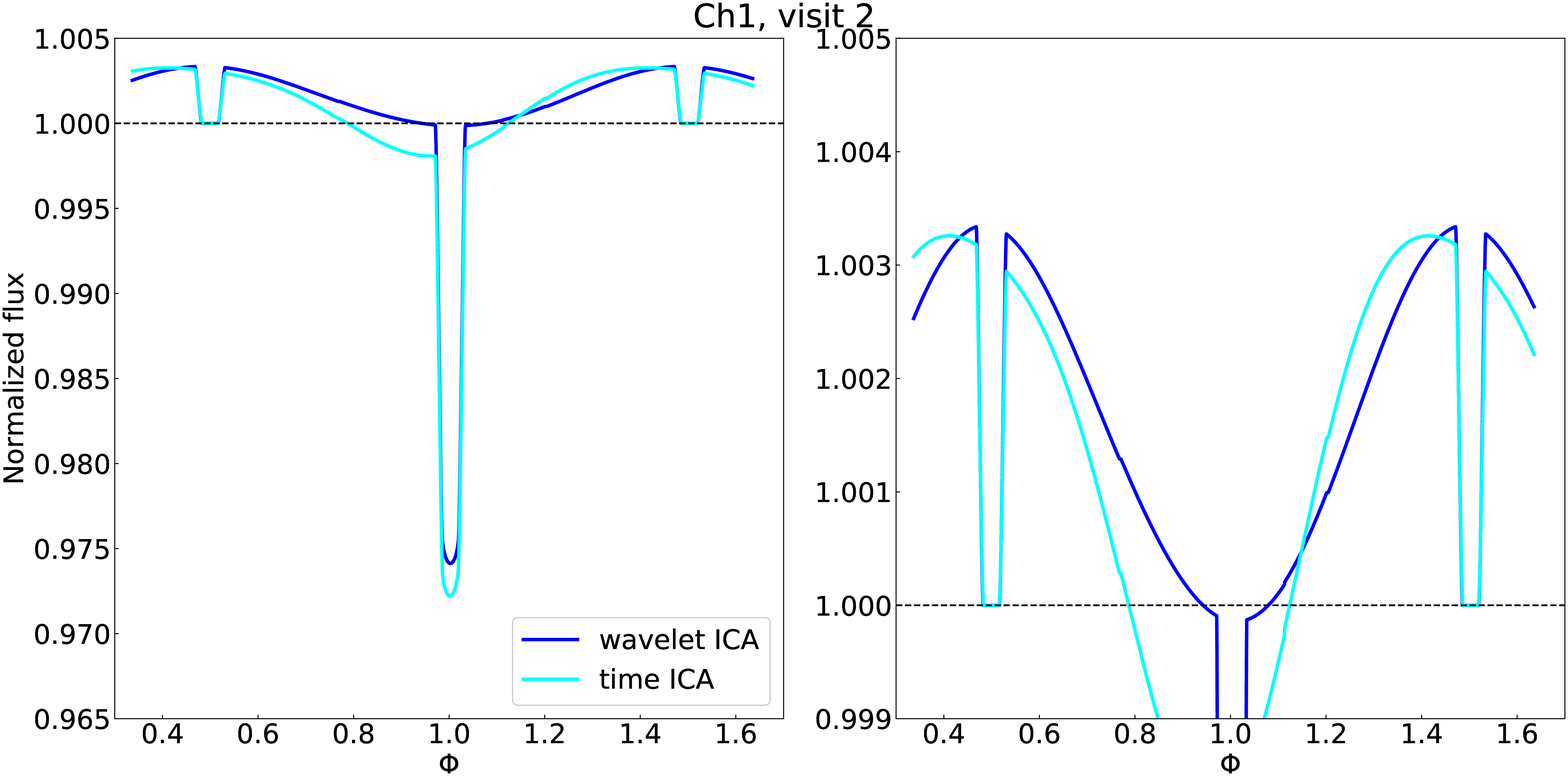}
\plotone{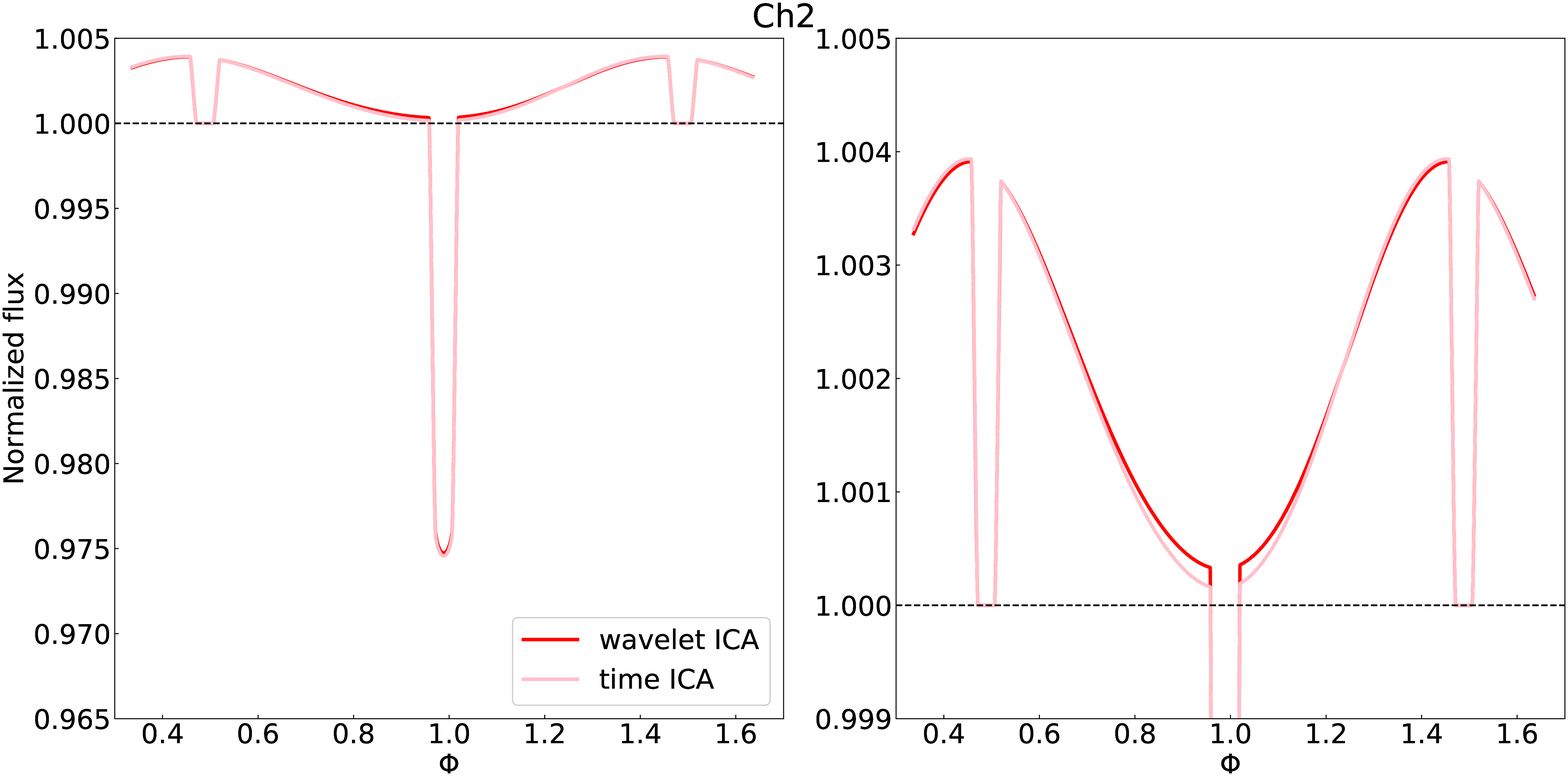}
\caption{Best-fit phase curve models obtained by using the wavelet pixel-ICA (darker colors) and time pixel-ICA techniques (lighter colors). The right panels are zoom of the left panels.
\label{fig:phasecurves_wavetime}}
\end{figure*}

\begin{figure*}[!h]
\epsscale{0.95}
\plotone{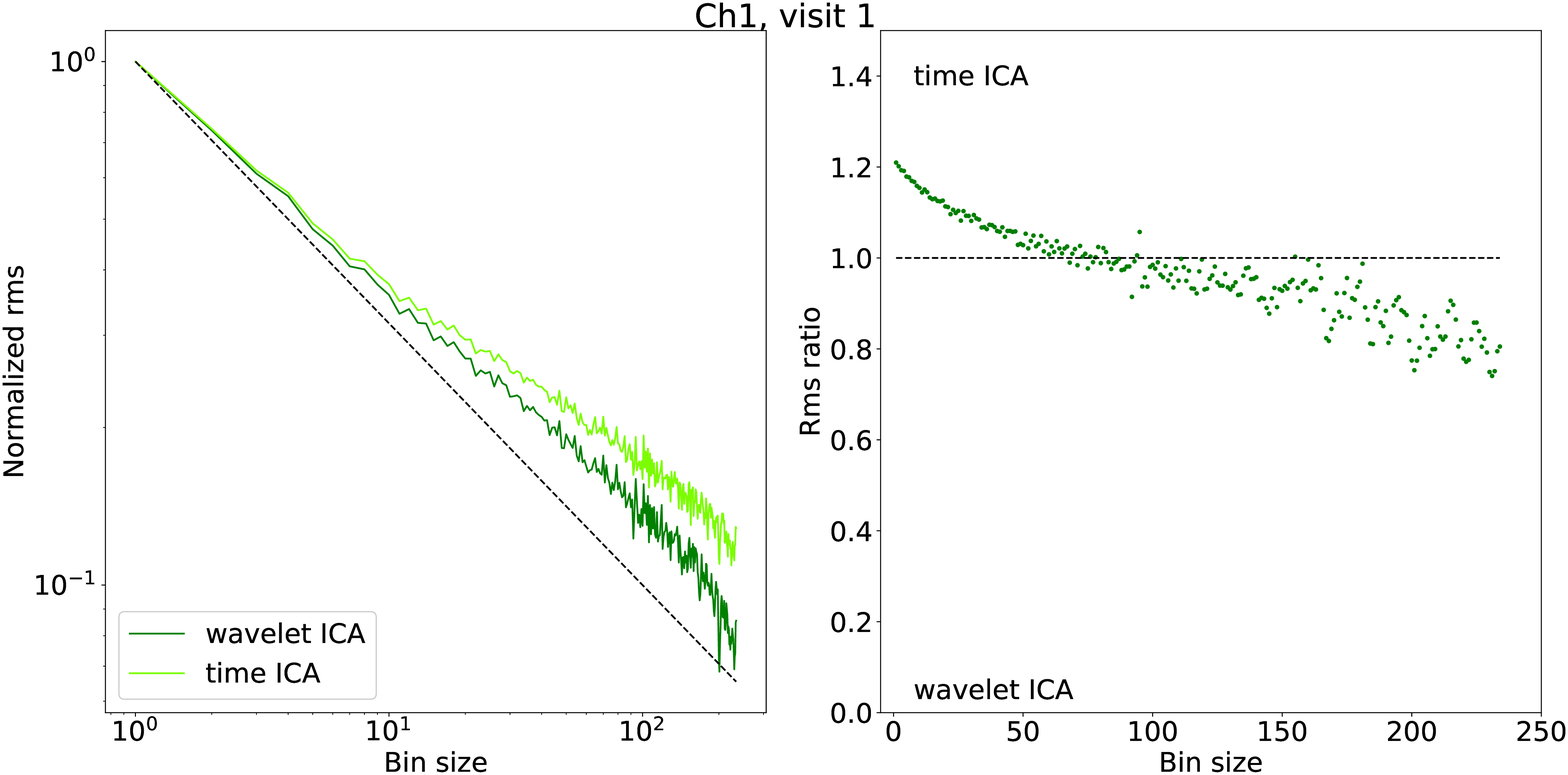}
\plotone{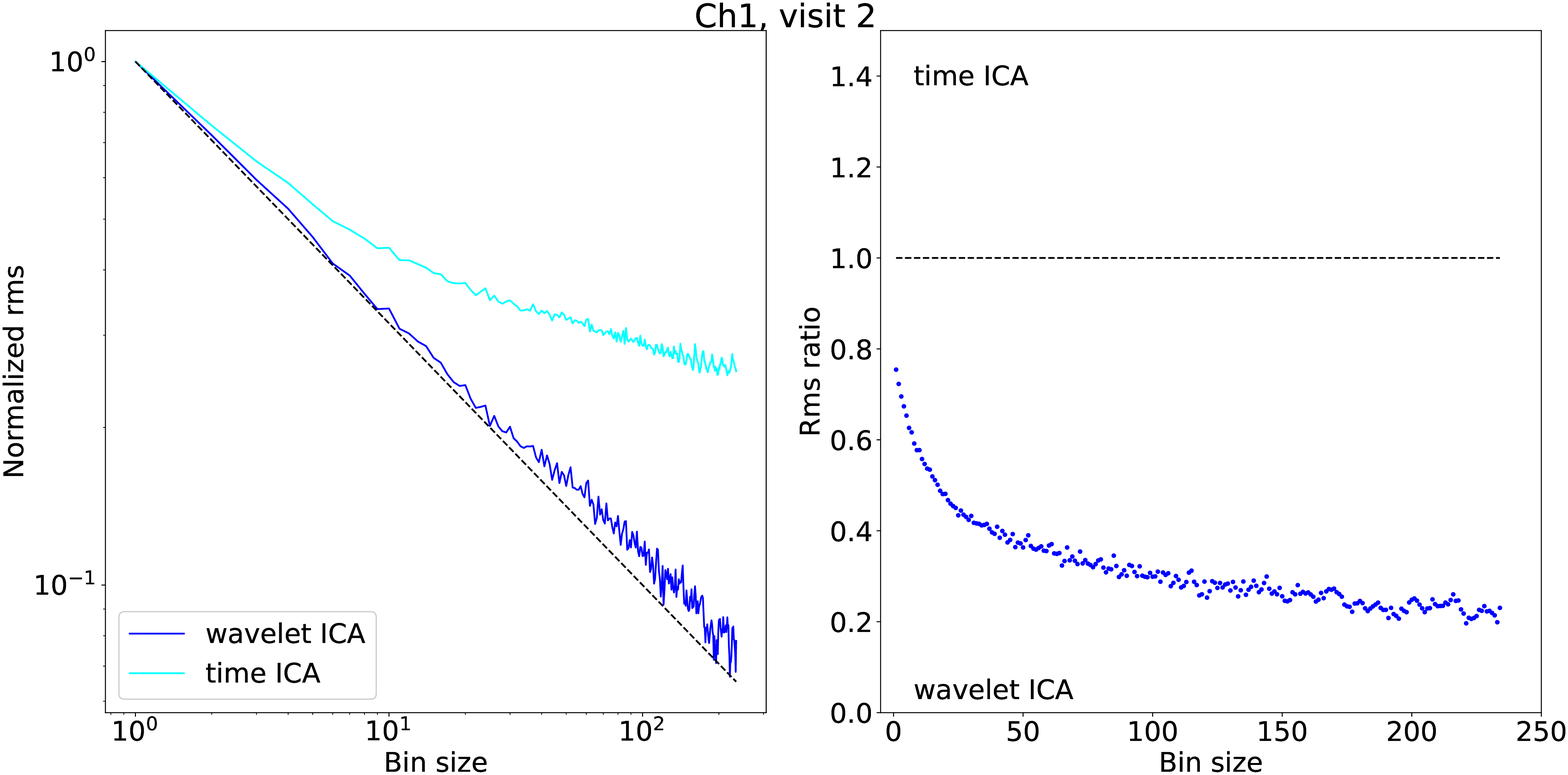}
\plotone{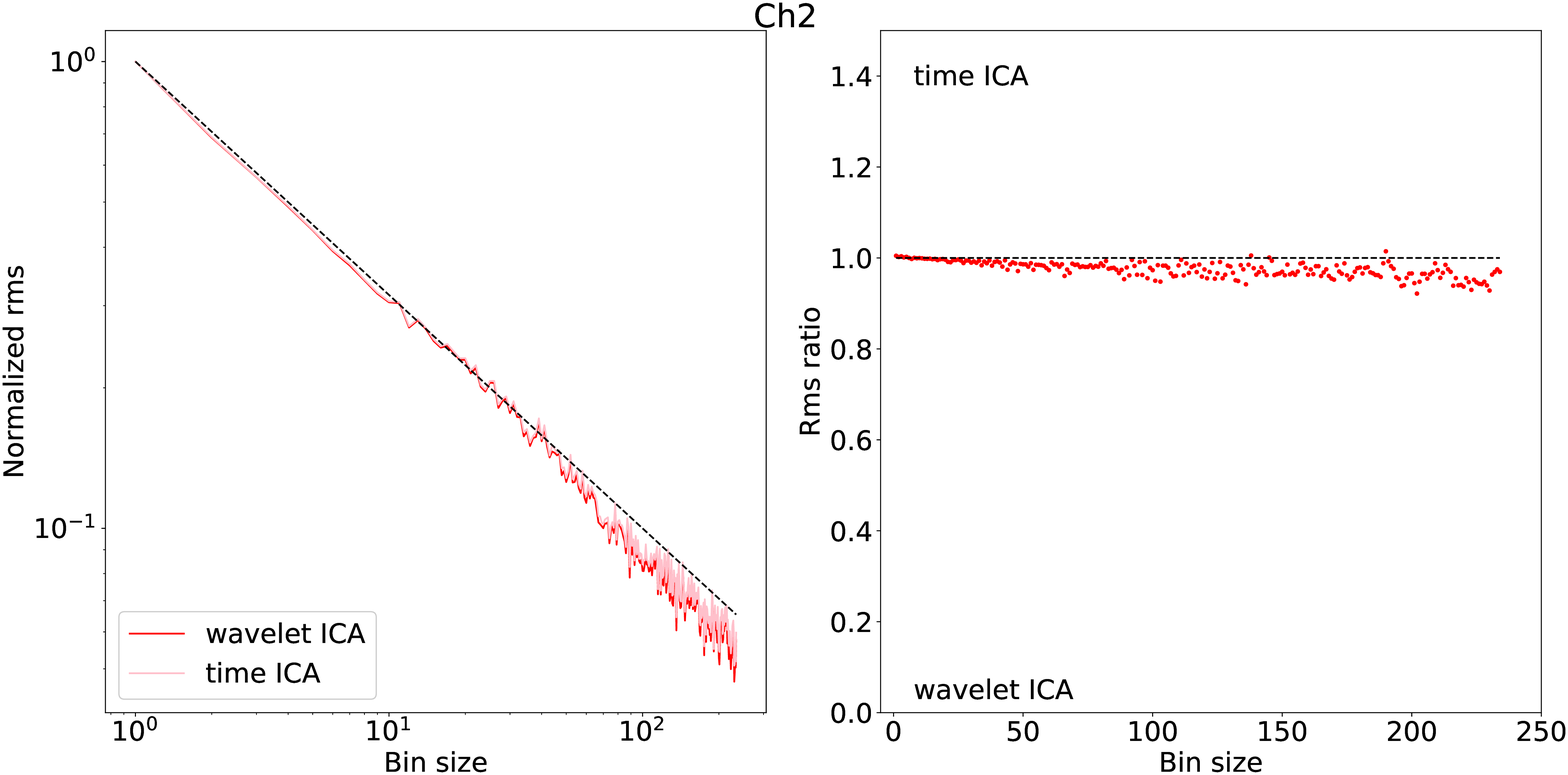}
\caption{Left panels: normalized rms of residuals as a function of bin size obtained by using the wavelet pixel-ICA (darker colors) and time pixel-ICA techniques (lighter colors). The black dashed lines show the theoretical behavior for gaussian residuals. Right panels: ratio between the rms of residuals obtained by using the wavelet pixel-ICA and time pixel-ICA techniques as a function of bin size. The black dashed lines denotes the separation (ratio = 1); the points below the lines correspond to the case of smaller residuals obtained by using the wavelet pixel-ICA (and vice versa).
\label{fig:binning_rmsres_scale_wavetime}}
\end{figure*}

Figure~\ref{fig:phasecurves_wavetime} compares the best-fit phase curve model obtained using pixel-ICA in the time or wavelet domains. The two approaches led to similar phase curve models for the 4.5~$\mu$m observation. Instead, the models obtained for the 3.6~$\mu$m visits using time pixel-ICA are less reliable, as they assume strong negative emission from the exoplanet nightside. Figure~\ref{fig:binning_rmsres_scale_wavetime} shows that the worse phase curve models are associated with higher levels of correlated noise in the fitting residuals, though, in some cases, the rms amplitudes are smaller. This study suggests that the alternative pixel-ICA algorithms are equivalent below a certain level of correlated noise (e.g., at 4.5~$\mu$m), otherwise the wavelet-based approach outperforms the analysis in the time domain.

For the transit-only analyses, the parameters obtained with the time pixel-ICA are consistent with those obtained with the wavelet pixel-ICA within less than 0.5~$\sigma$ (see Figure~\ref{fig:p2bT0_phase_tran_A100}). It is reasonable to expect that the impact of low-frequency noise is smaller over the transit timescale, therefore explaining the apparent equivalence of the two methods. 

For the eclipse-only analyses, the eclipse depths obtained with the time pixel-ICA are systematically larger, and with larger error bars, than those obtained with the wavelet pixel-ICA. In some cases, the eclipse depth estimates obtained with the time pixel-ICA are above their phase curve upper limits. Therefore, the wavelet pixel-ICA outperforms the time pixel-ICA in eclipse-only analyses. This fact was already observed in the previous literature \citep{morello16}, and it is attributed to the smaller signal-to-noise ratio of the eclipse signal. For the 4.5~$\mu$m eclipses (least correlated noise), the two methods lead to consistent results within 1~$\sigma$.

\begin{table}[!t]
\begin{center}
\begin{threeparttable}
\caption{$\Delta$BIC and $\Delta$AIC for the different long-term ramp models and types of observation. \label{tabBICAIC}}
\begin{tabular}{cc|cc|cc|cc}
\tableline\tableline
 &  & \multicolumn{2}{|c|}{Ch2} & \multicolumn{2}{|c|}{Ch1, Visit 2}  & \multicolumn{2}{|c}{Ch1, Visit 1} \\
Obs. type & Ramp model & $\Delta$BIC & $\Delta$AIC  & $\Delta$BIC & $\Delta$AIC  & $\Delta$BIC & $\Delta$AIC \\
\tableline
\multirow{3}{*}{full} & Constant & \textbf{0.0} & 0.0 & +310.8 & +324.0 & \textbf{0.0} & +10.7 \\
 & Linear & +7.0 & +0.4 & +8.9 & +15.6 & +1.7 & +5.7 \\
 & Quadratic & +13.2 & 0.0 & \textbf{0.0} & 0.0 & +2.6 & 0.0 \\
\tableline
\multirow{3}{*}{ecl1 + tr} & Constant & \textbf{0.0} & 0.0 & +2.2 & +8.3 & \textbf{0.0} & +6.1 \\
 & Linear & +6.4 & +0.1 & \textbf{0.0} & 0.0 & +0.1 & 0.0 \\
 & Quadratic & +13.0 & +0.5 & +7.3 & +1.2 & +8.2 & +1.8 \\
\tableline
\multirow{3}{*}{tr + ecl2} & Constant & \textbf{0.0} & 0.0 & \textbf{0.0} & 4.1 & \textbf{0.0} & +9.6 \\
 & Linear & +7.8 & +1.6 & +2.1 & 0.0 & +5.8 & +9.2 \\
 & Quadratic & +16.0 & +3.5 & +9.0 & +0.7 & +2.8 & 0.0 \\
\tableline
\end{tabular}
\begin{tablenotes}
\item ``full'' = full phase curve; ``ecl1 + tr'' = half phase curve including the eclipse prior transit; ``tr + ecl2'' = half phase curve including the eclipse after transit.
\end{tablenotes}
\end{threeparttable}
\end{center}
\end{table}

\section{Half phase curves}
\label{app_halfpc}
We discuss here the analysis of the so-called ``half phase curves'', i.e., continuous observations including one transit and either the eclipse immediately before or after. In this study, we obtain two half phase curves from each visit by considering two out of three consecutive AORs (where applicable). For the second 3.6~$\mu$m visit, we split the first AOR in two parts in order to get three pieces as in the other visits. Note that the half phase curves within the same visit share the same transit event.

\begin{figure*}[!t]
\epsscale{0.95}
\plotone{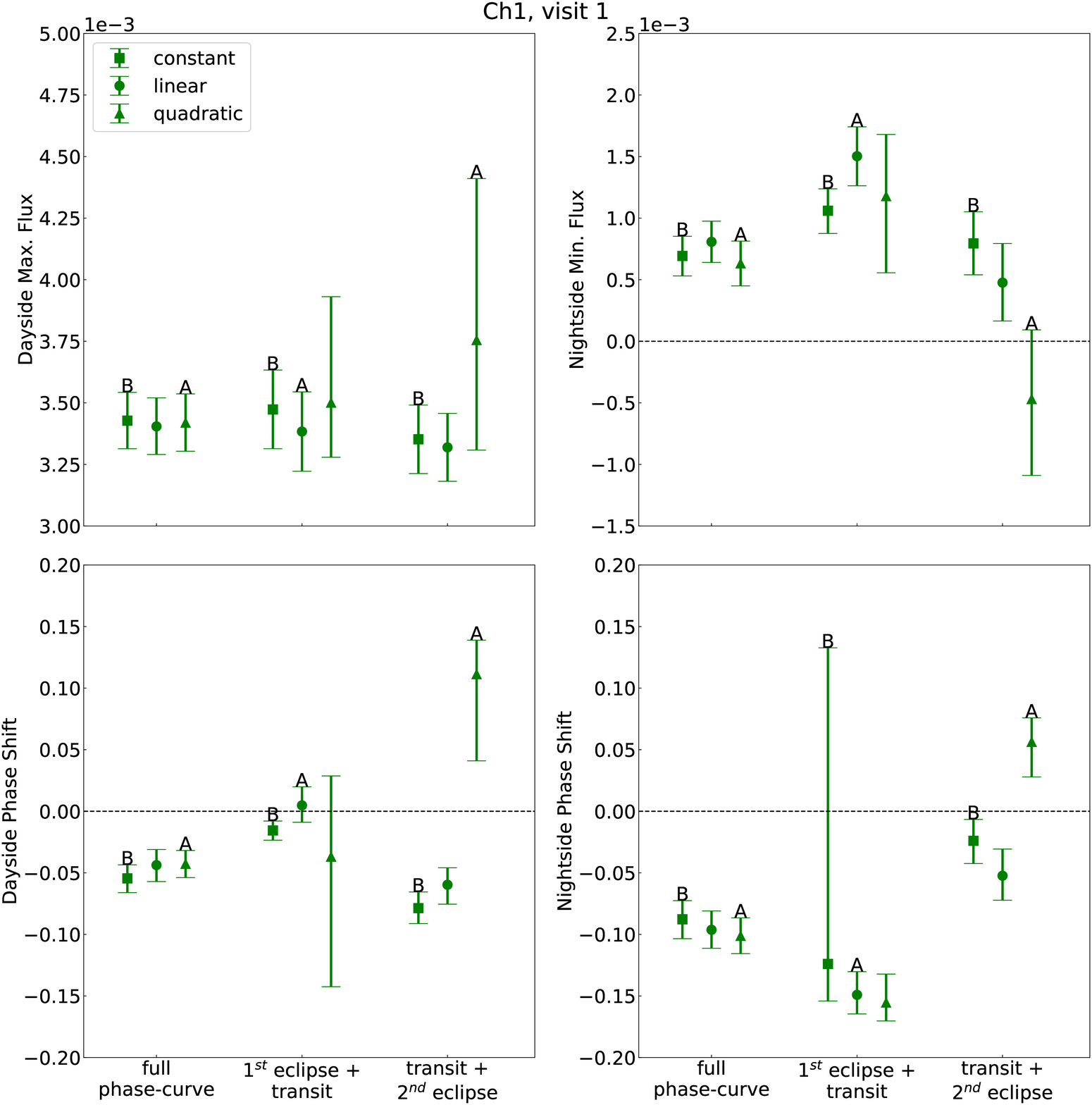}
\caption{Top, left panel: maximum exoplanetary flux, relative to the stellar flux, for the first 3.6~$\mu$m visit from the full and half phase curve analyses by using the different ramp models (see Appendix~\ref{app_halfpc}). The letters ``A'' and ``B'' indicate the minimum AIC and BIC solutions among the different ramp models. Top, right panel: analogous plot for the minimum exoplanetary flux. Bottom, left panel: analogous plot for the offset of the phase curve maximum relative to mid-eclipse. Bottom, right panel: analogous plot for the offset of the phase curve minimum relative to mid-transit.
\label{fig:daynight_r5236_ch1_allcases}}
\end{figure*}

\begin{figure*}[!t]
\epsscale{0.95}
\plotone{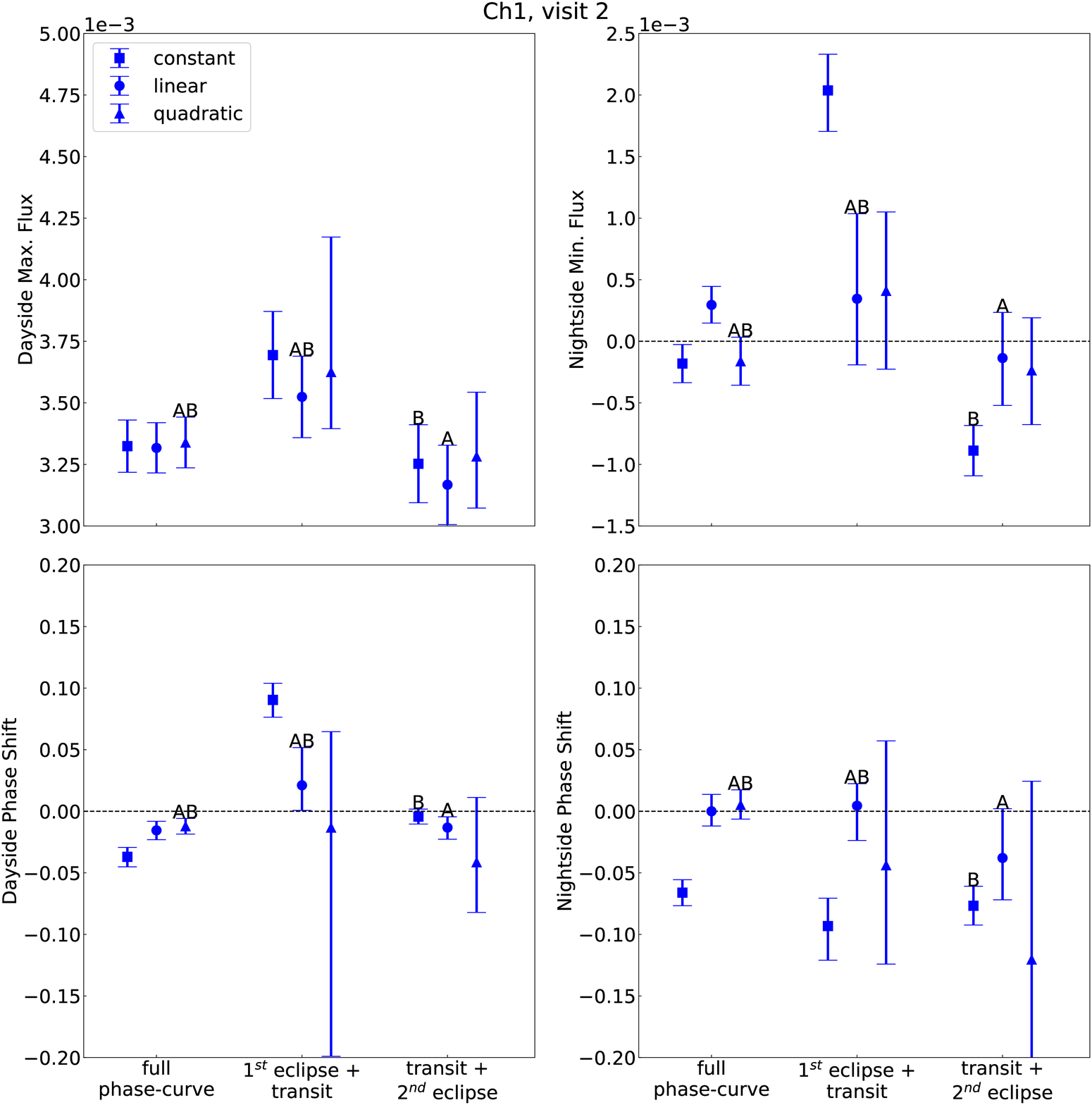}
\caption{Top, left panel: maximum exoplanetary flux, relative to the stellar flux, for the second 3.6~$\mu$m visit from the full and half phase curve analyses by using the different ramp models (see Appendix~\ref{app_halfpc}). The letters ``A'' and ``B'' indicate the minimum AIC and BIC solutions among the different ramp models. Top, right panel: analogous plot for the minimum exoplanetary flux. Bottom, left panel: analogous plot for the offset of the phase curve maximum relative to mid-eclipse. Bottom, right panel: analogous plot for the offset of the phase curve minimum relative to mid-transit.
\label{fig:daynight_r57744_ch1_allcases}}
\end{figure*}

\begin{figure*}[!t]
\epsscale{0.95}
\plotone{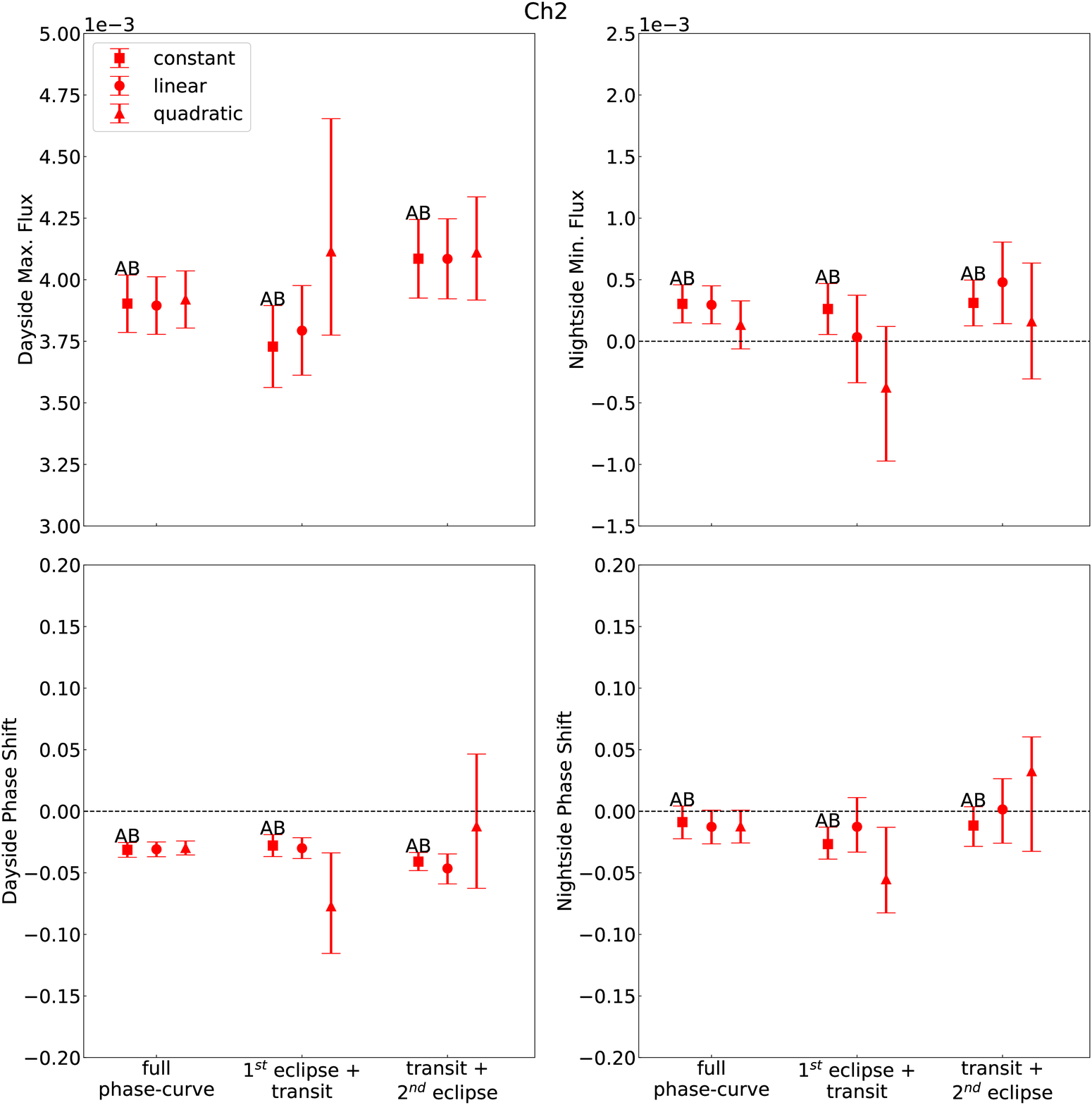}
\caption{Top, left panel: maximum exoplanetary flux, relative to the stellar flux, for the f4.5~$\mu$m visit from the full and half phase curve analyses by using the different ramp models (see Appendix~\ref{app_halfpc}). The letters ``A'' and ``B'' indicate the minimum AIC and BIC solutions among the different ramp models. Top, right panel: analogous plot for the minimum exoplanetary flux. Bottom, left panel: Analogous plot for the offset of the phase curve maximum relative to mid-eclipse. Bottom, right panel: analogous plot for the offset of the phase curve minimum relative to mid-transit.
\label{fig:daynight_r51777_ch2_allcases}}
\end{figure*}

\begin{figure}[!ht]
\epsscale{0.65}
\plotone{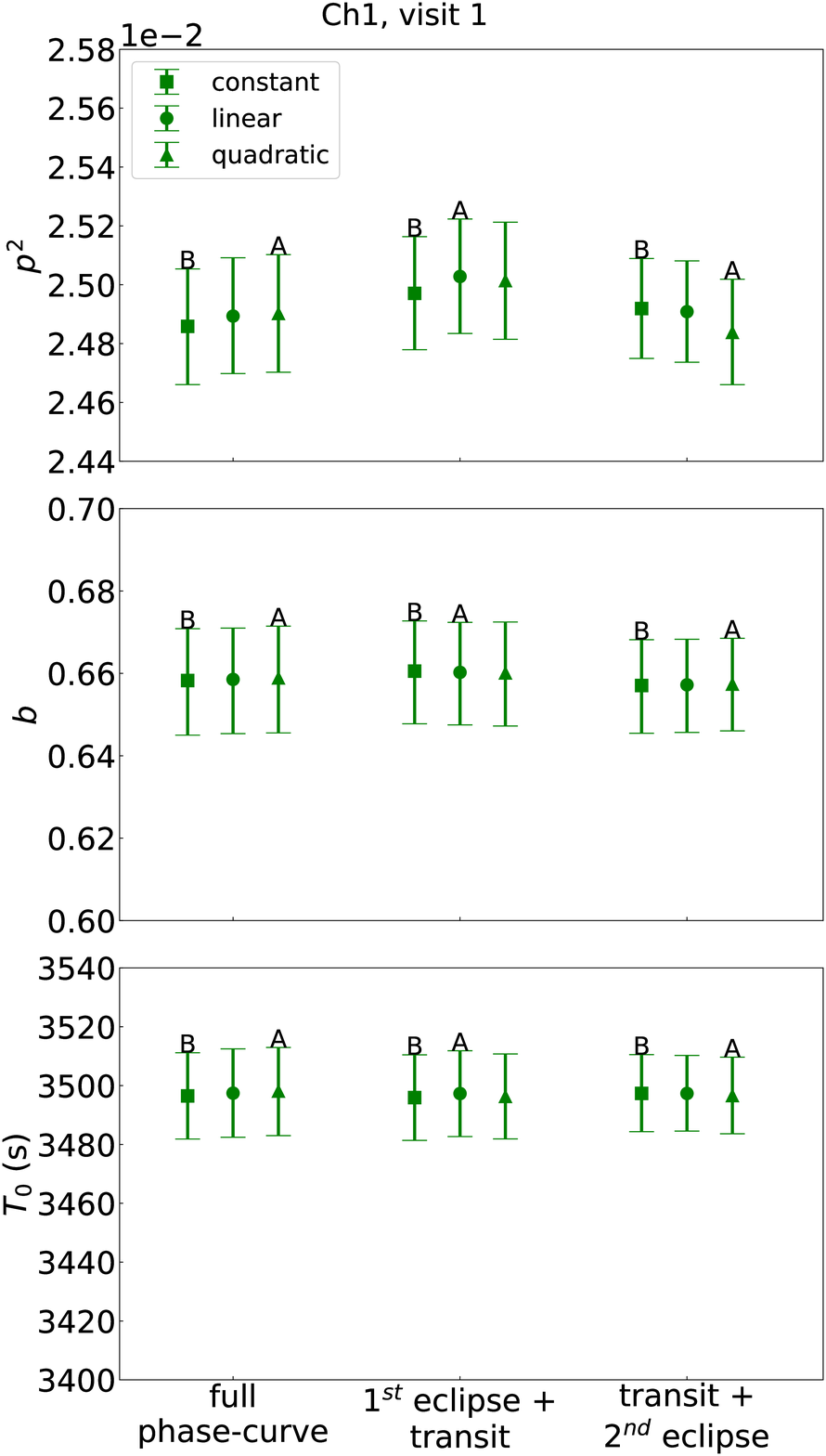}
\caption{Top panel: transit depth estimates for the first 3.6~$\mu$m visit from the full and half phase curve analyses by using the different ramp models (see Appendix~\ref{app_halfpc}). The letters ``A'' and ``B'' indicate the minimum AIC and BIC solutions among the different ramp models. Middle and bottom panels: analogous plots for the impact parameter and for the transit duration.
\label{fig:p2bT0_r5236_ch1_allcases}}
\end{figure}

\begin{figure}[!ht]
\epsscale{0.65}
\plotone{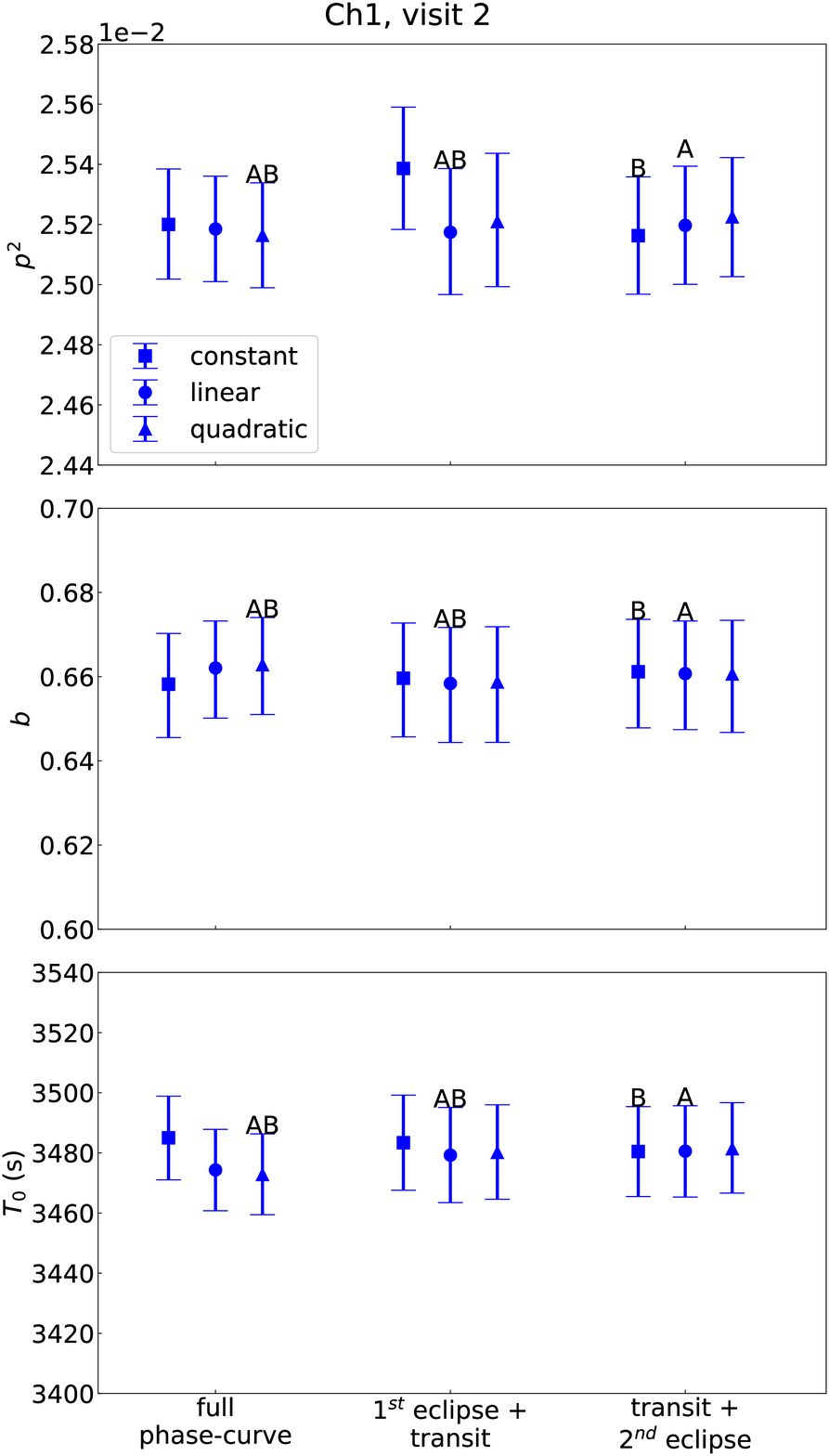}
\caption{Top panel: transit depth estimates for the second 3.6~$\mu$m visit from the full and half phase curve analyses by using the different ramp models (see Appendix~\ref{app_halfpc}). The letters ``A'' and ``B'' indicate the minimum AIC and BIC solutions among the different ramp models. Middle and bottom panels: analogous plots for the impact parameter and for the transit duration.
\label{fig:p2bT0_r57744_ch1_allcases}}
\end{figure}

\begin{figure}[!ht]
\epsscale{0.65}
\plotone{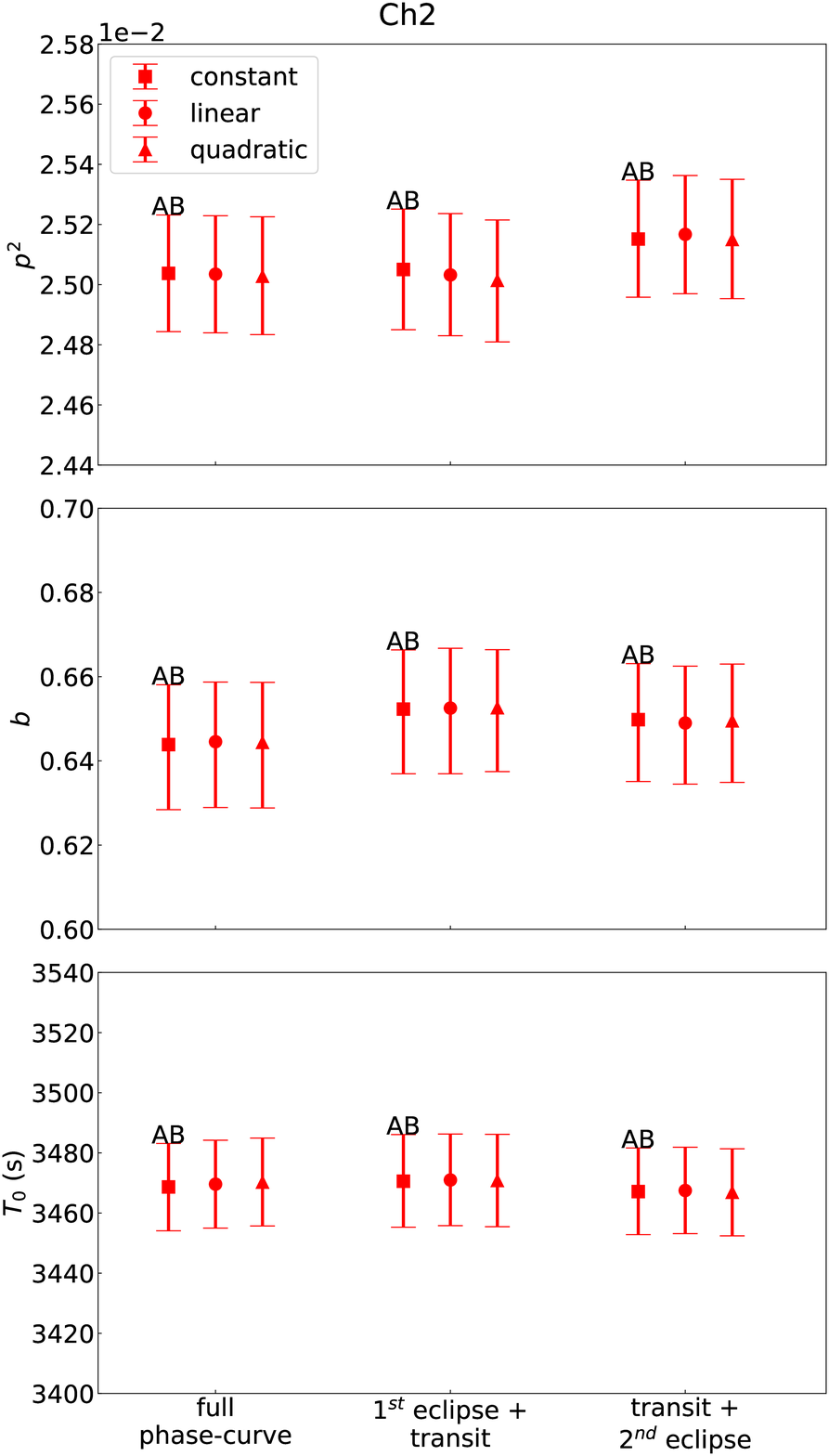}
\caption{Top panel: transit depth estimates for the 4.5~$\mu$m visit from the full and half phase curve analyses by using the different ramp models (see Appendix~\ref{app_halfpc}). The letters ``A'' and ``B'' indicate the minimum AIC and BIC solutions among the different ramp models. Middle and bottom panels: analogous plots for the impact parameter and for the transit duration.
\label{fig:p2bT0_r51777_ch2_allcases}}
\end{figure}

Figures~\ref{fig:daynight_r5236_ch1_allcases}-\ref{fig:daynight_r51777_ch2_allcases} report the phase curve parameters for the full and half phase curve analyses with a constant, linear, and quadratic ramp model. Figures~\ref{fig:p2bT0_r5236_ch1_allcases}-\ref{fig:p2bT0_r51777_ch2_allcases} report the corresponding transit parameters. Table~\ref{tabBICAIC} reports the relative $\Delta$BIC and $\Delta$AIC for the different ramp models.

\begin{figure}[!ht]
\epsscale{0.95}
\plotone{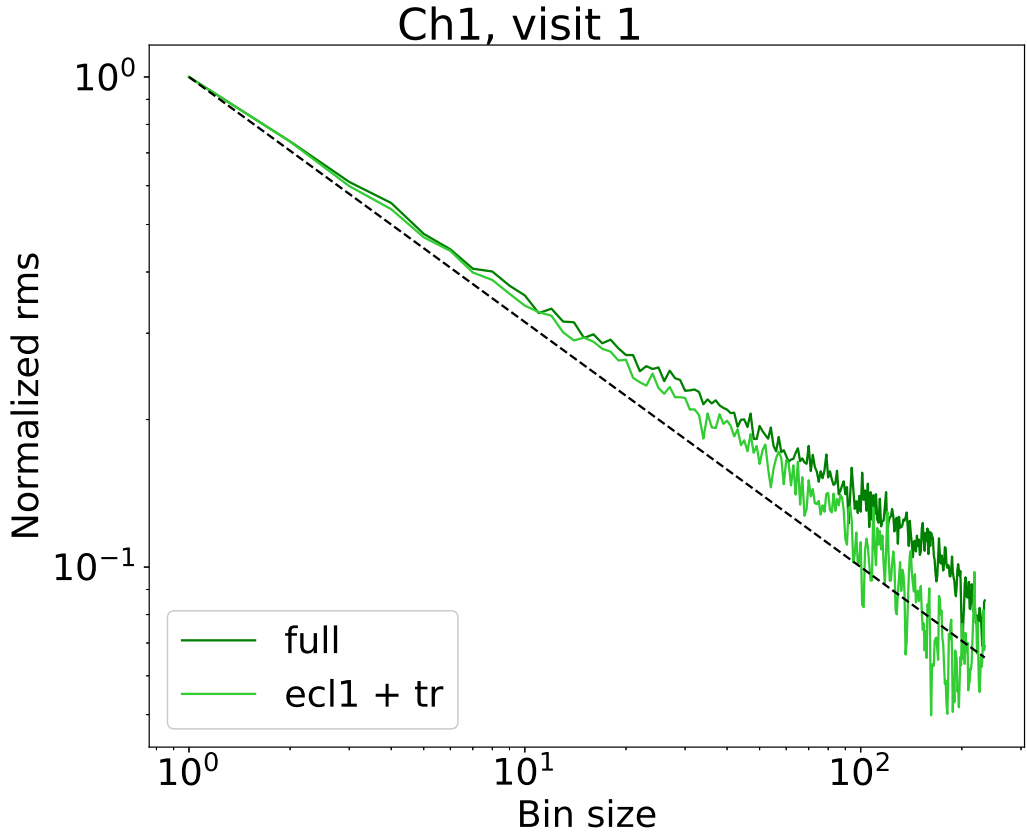}
\caption{Normalized rms of residuals as a function of the bin size for the first 3.6~$\mu$m visit. The full phase curve analysis is represented as the dark green line and the half phase curve analysis including the eclipse prior as the light green line. The black dashed line shows the theoretical behavior for Gaussian residuals.
\label{fig:binning_rmsres_scale_ch1v1_special}}
\end{figure}

The phase curve parameters are more degenerate with the ramp parameters in the half phase curve models, as suggested by the much larger and asymmetric error bars. The transit parameters are independent on the choice of ramp model, half or full phase curve, i.e., their dispersion are much smaller than their error bars.
Only for the first 3.6~$\mu$m visit, the analysis of the half phase curve, including the eclipse before the transit, outperforms the full phase curve analysis. Figure~\ref{fig:binning_rmsres_scale_ch1v1_special} shows that the correlated noise in the half phase curve residuals is significantly smaller than in the full phase curve residuals. Also, the dayside shift resulting from the half phase curve is in better agreement with the results from the other observations.



\section{Limb-darkening coefficients}
\label{app_ldc}

\begin{figure}[!ht]
\epsscale{0.9}
\plotone{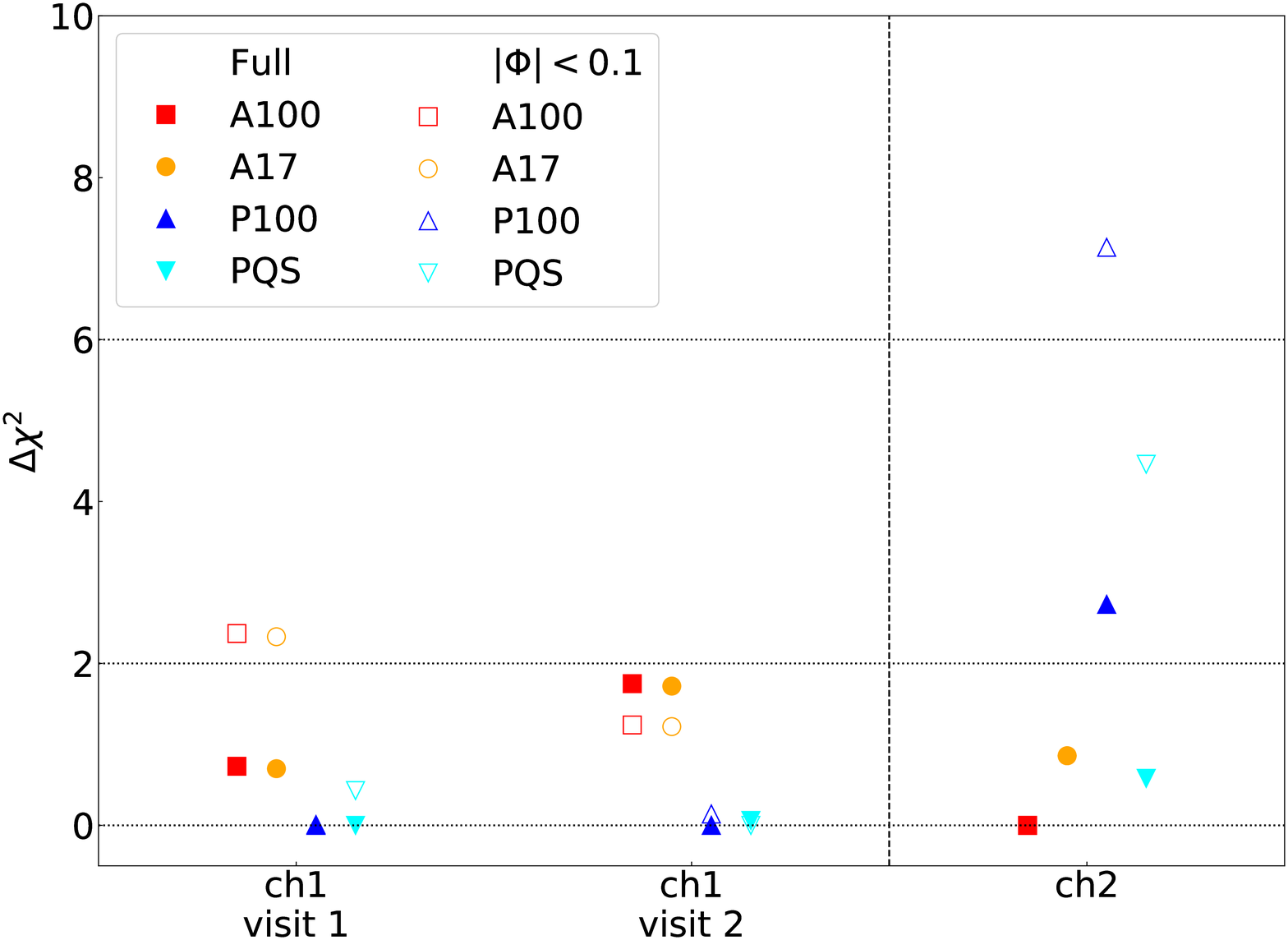}
\caption{Relative chi-square obtained by using different sets of limb-darkening coefficients: A100 (red squares), A17 (orange circles), P100 (blue, upward triangles), and PQS (cyan, downward triangles). The full markers refer to the full phase curve residuals. The empty markers refer to the sub-interval of the same residuals centered on the transit. The horizontal lines delimit the significance levels according to \cite{raftery95}: $\Delta \chi^2 \le$2 is not significant, 2$< \Delta \chi^2 \le$6 denotes positive evidence against the model with higher $\chi^2$, and 6$< \Delta \chi^2 \le$10 denotes strong evidence against the model with higher $\chi^2$.
\label{fig:ld_deltabic}}
\end{figure}
Figure~\ref{fig:ld_deltabic} shows the $\chi^2$ differences between the light curve fits with the different sets of limb-darkening coefficients. In all cases, the $\chi^2$ differences are smaller than 2, except the 2.7 difference between the P100 and A100 models of the 4.5~$\mu$m light curve. Such differences are not significant or barely significant according to \cite{raftery95}. Given that the limb-darkening coefficients only affect the points during the transit, we recalculated the $\chi^2$ differences over the phase interval $| \Phi | \le $0.1. The new differences only indicate a strong evidence ($\Delta \chi^2 = $7.1) against the P100 model, and a positive evidence ($\Delta \chi^2 = $4.5) against the PQS model at 4.5~$\mu$m. 

\begin{figure}[!ht]
\epsscale{0.9}
\plotone{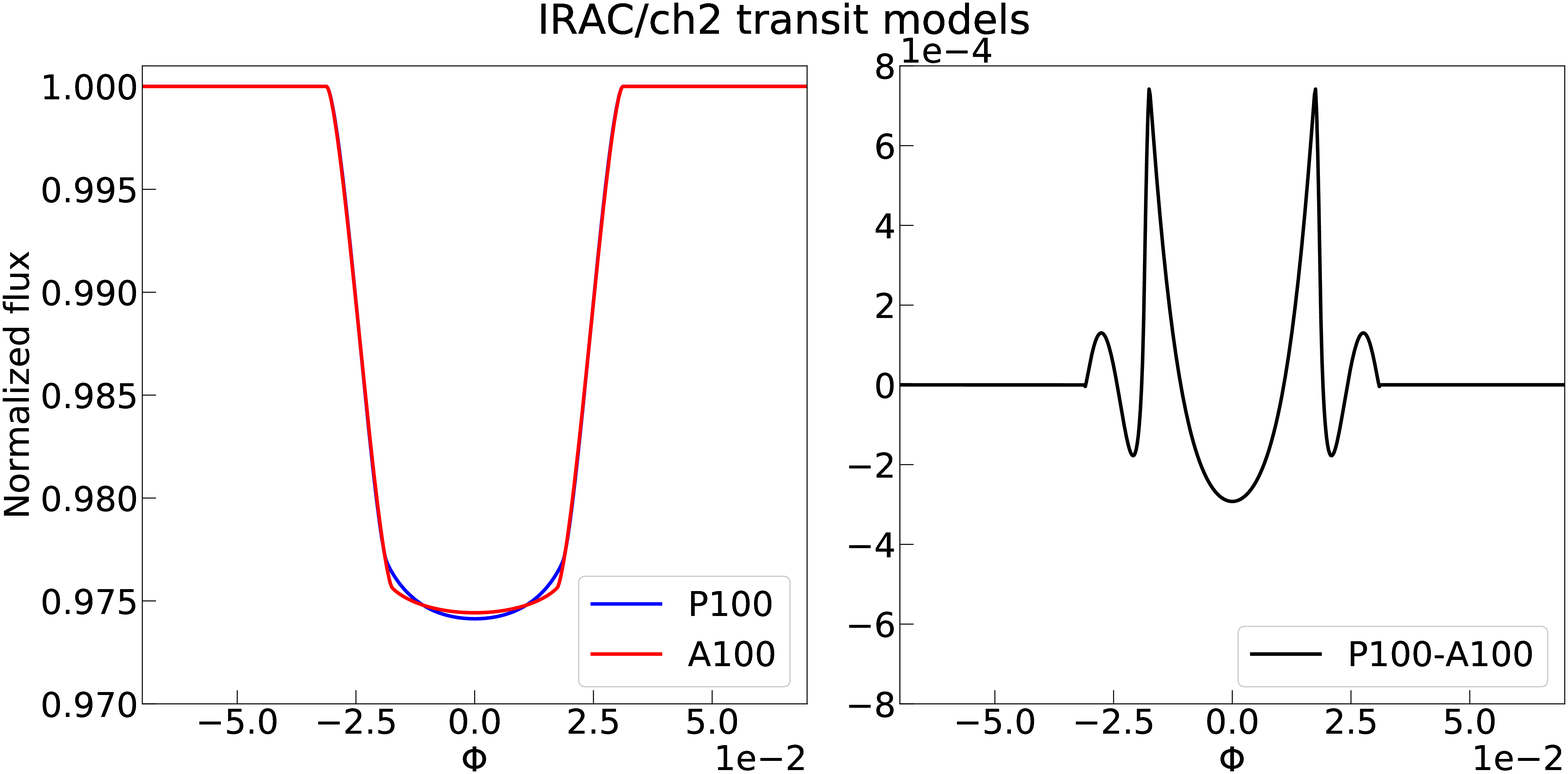}
\caption{Left panel: best-fit transit models for the 4.5~$\mu$m visit obtained by using A100 (red) and P100 (blue) limb-darkening coefficients. Right panel: difference between the alternative transit models.
\label{fig:transitmodels_A100vsP100_ch2}}
\end{figure}

\begin{figure}[!ht]
\epsscale{0.9}
\plotone{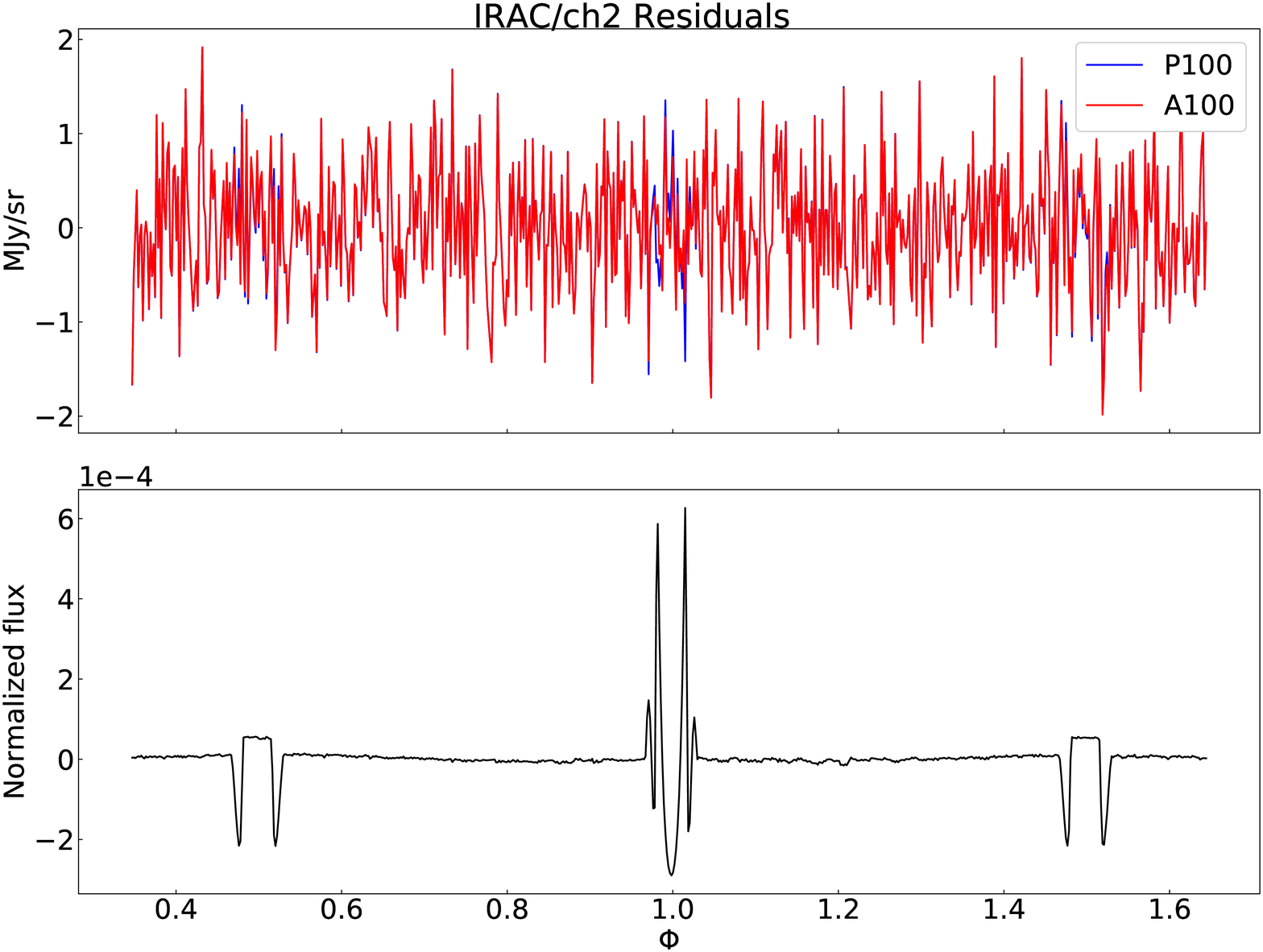}
\caption{Top panel: light curve residuals of the 4.5~$\mu$m visit obtained by using A100 (red) and P100 (blue) limb-darkening coefficients. Bottom panel: difference between the residual time series above. Note that the only differences occur during the transit and eclipses. The difference between the residuals obtained with different limb-darkening coefficients is smaller for the other observations.
\label{fig:res_A100vsP100_ch2}}
\end{figure}

Figure~\ref{fig:transitmodels_A100vsP100_ch2} compares the two transit models obtained with the P100 and A100 limb-darkening coefficients at 4.5~$\mu$m, that led to the largest $\Delta \chi^2$ in the light curve residuals.
Figure~\ref{fig:res_A100vsP100_ch2} shows the difference between the corresponding light curve residuals. The difference is non-zero only during the transit and the two eclipses\footnote{Even if, the stellar limb-darkening does not affect the eclipse shape, the eclipse duration is affected, as it is equal to the transit duration.} and the maximum peaks are $\sim$600 ppm. The rms amplitude of the residuals is 1870 ppm, i.e., more than three times larger than the maximum difference.





\clearpage

\clearpage

\end{document}